\documentclass[a4paper,UKenglish,cleveref, autoref, thm-restate]{lipics-v2021}

\pdfoutput=1 
\hideLIPIcs  

\usepackage[export]{adjustbox}
\usepackage{subcaption}
\usepackage{tikz}
\usepackage{lipsum}
\usepackage[export]{adjustbox}
\usepackage{textcomp}
\usepackage{supertabular}
\usepackage{enumerate}
\usepackage{booktabs}
\usepackage{pifont}
\usepackage{url}

\usepackage{flushend}

\newcommand\T[1]{\vspace{4pt}\textbf{#1}}
\newcommand\TA[1]{\vspace{0pt}\textbf{#1}}

\title{DeFi Lending During The Merge} 

\author{Lioba Heimbach}{ETH Zurich, Switzerland}{hlioba@ethz.ch}{0000-0002-8258-1712}{}
\author{Eric Schertenleib}{ETH Zurich, Switzerland}{ericsch@ethz.ch}{0000-0002-0927-8178}{}
\author{Roger Wattenhofer}{ETH Zurich, Switzerland}{wattenhofer@ethz.ch}{0000-0002-6339-3134}{}

\authorrunning{L. Heimbach, E. Schertenleib and R. Wattenhofer} 

\Copyright{Lioba Heimbach, Eric Schertenleib and Roger Wattenhofer} 


\category{} 

\relatedversion{} 

\nolinenumbers 

\EventEditors{Joseph Bonneau and S. Matthew Weinberg}
\EventNoEds{2}
\EventLongTitle{5th Conference on Advances in Financial Technologies (AFT 2023)}
\EventShortTitle{AFT 2023}
\EventAcronym{AFT}
\EventYear{2023}
\EventDate{October 23-25, 2023}
\EventLocation{Princeton, NJ, USA}
\EventLogo{}
\SeriesVolume{282}
\ArticleNo{9}

\begin{document}

\maketitle

\begin{abstract}
Lending protocols in decentralized finance enable the permissionless exchange of capital from lenders to borrowers without relying on a trusted third party for clearing or market-making. Interest rates are set by the supply and demand of capital according to a pre-defined function. In the lead-up to \textit{The Merge}: Ethereum blockchain's transition from \textit{proof-of-work (PoW)} to \textit{proof-of-stake (PoS)}, a fraction of the Ethereum ecosystem announced plans of continuing with a PoW-chain. Owners of ETH -- whether their ETH was borrowed or not -- would hold the native tokens on each chain. This development alarmed lending protocols. They feared spiking ETH borrowing rates would lead to mass liquidations which could undermine their viability. Thus, the decentralized autonomous organization running the protocols saw no alternative to intervention -- restricting users' ability to borrow. 

We investigate the effects of the merge and the aforementioned intervention on the two biggest lending protocols on Ethereum: AAVE and Compound. Our analysis finds that borrowing rates were extremely volatile, jumping by two orders of magnitude, and borrowing at times reached 100\% of the available funds. Despite this, no spike in mass liquidations or irretrievable loans materialized. Further, we are the first to quantify and analyze \textit{hard-fork-arbitrage}, profiting from holding debt in the native blockchain token during a hard fork. We find that arbitrageurs transferred tokens to centralized exchanges which at the time were worth more than 13~Mio~US\$, money that was effectively extracted from the platforms' lenders. 
\end{abstract}

\section{Introduction}\label{sec:intro}

\newcommand{\cmark}{\ding{51}}%
\newcommand{\xmark}{\ding{55}}%

Participants in financial markets borrow or lend for a plethora of reasons: raising capital for investments such as buying a house, saving for retirement, or short-selling securities. Particularly consequential for the stability of the financial market is leveraged trading. Thereby, a trader takes on leverage by borrowing from a counterparty, typically a broker or bank, to buy financial securities. The lender must approve the borrowing and puts certain restrictions in place to safeguard their funds (margin requirements). Given the counterparty risks, i.e., the risk that the other party defaults on their contractual obligation, regulators closely monitor these activities and enforce certain restrictions to reduce the likelihood of major financial upheaval. 

\emph{Decentralized Finance (DeFi)} promises to offer financial services to users without requiring prior clearance or a known and trustworthy counterparty. Instead, DeFi is built using \emph{smart contracts}, i.e., executable code on the blockchain. Lending protocols have a central role in the DeFi protocol space. They allow anyone to become a lender by depositing their funds in the protocol. Further, anyone can borrow funds as long as their deposited assets exceed their borrowed funds in value by a pre-defined margin (over-collateralization). The restriction to over-collateralize loans on lending protocols is their key to unlocking trustless loans. Additionally, the over-collateralization margin aims to ensure that there is enough time for a position at risk of becoming under-collateralized to become liquidated in time for the debt to be recovered. 

Thus, it is crucial for lending protocols to correctly assess the risks of the crypto assets they allow as collateral in the market and to set the margin accordingly. Lending protocols strive to find the right balance between (i) offering competitive rates to borrowers and (ii) low risks for lenders. As the risks associated with the various collateral assets are likely to change over time, lending protocols can adjust various risk-related parameters, such as the over-collateralization margin. These changes are generally discussed and decided by the \emph{decentralized autonomous organization (DAO)} that governs the protocol. 

The Ethereum blockchain is the birthplace of DeFi and the home of the leading lending protocols in terms of total value locked~\cite{2023lending}. Initially, the consensus mechanism of the Ethereum blockchain relied on the energy-intensive \emph{proof-of-work (PoW)}. Ethereum had planned for years to switch to the more energy-efficient \emph{proof-of-stake (PoS)}. \emph{The Merge}, which was executed in September 2022, marked the end of PoW and the start of PoS. However, in the lead-up to the merge, there was opposition from parts of the Ethereum mining community whose business model relied on PoW. They pushed for a hard fork that would retain a PoW blockchain. Thus, even though the vast majority of the Ethereum community announced that they would switch to PoS, it was unclear how the value of ETH would be distributed between the two blockchains. 

The approaching merge and possibility of a hard fork, as well as the resulting distribution of value between the two blockchains posed a challenge to lending protocols. Anyone who held ETH on the blockchain in the last block before the merge would then own the same number of tokens on both the PoS chain and the PoW chain after the merge. Thus, given that future markets indicated that the PoW token (ETHW) would retain a value of a few percentage points of its PoS counterpart, some users borrowed ETH ahead of the merge in order to receive both tokens after the merge. We will refer to this as \emph{hard-fork-arbitrage}, a form of event-driven arbitrage. 

Thus, protocols expected the demand for ETH borrowing to increase drastically and, as a result, expected both ETH borrowing rates and the utilization, the ratio between the ETH loans and ETH liquidity, of the ETH market to skyrocket. Furthermore, given the intertwined nature of DeFi protocols and the central position lending protocols occupy therein, the stability of these platforms is essential.

In response, the DAOs governing AAVE and Compound, the two largest lending protocols on the Ethereum blockchain, decided to intervene. While AAVE paused ETH borrowing entirely ahead of the merge, Compound adjusted the risk parameters that determine the borrowing rate and capped ETH borrowing. This intervention, however, led to market distortions that adversely affected lenders, who were effectively on the losing side of the hard-fork-arbitrage. Furthermore, the lending market dried up as protocols ceased to have any available liquidity.

\subsection*{Our Contributions}
We analyze lending protocols during a critical time period. The execution of the merge on the Ethereum blockchain tested the resiliency of lending protocols toward significant external events. Thus, it provides a crucial case study of challenges faced by DeFi protocols under extraordinary circumstances. We analyze the effects of the merge on lending protocols as well as their attempts to mitigate borrowing rate spikes. They feared that such rate spikes would cause both mass liquidations, i.e., enforced repayment of the debt, and the accumulation of bad debt, i.e., debt that is not over-collateralized, leading to losses for lenders and potentially harming their own viability. Our analysis focuses on AAVE and Compound, the biggest DeFi lending protocols. Together AAVE and Compound account for more than 85\% of the volume locked on lending protocols on the Ethereum blockchain, the home of most DeFi applications~\cite{2023lending}. 
\begin{itemize}
    \itemsep0em 
    \item We find that the changes the protocols implemented, capping borrowing, may have helped prevent mass liquidations and the accumulation of bad debt but were only partially successful in keeping rates at normal levels.
    \item  We show that the protocols failed to adequately compensate their lenders, thus placing them at the losing end. The beneficiaries were arbitrageurs, including the now infamous Alameda research, who extracted in excess of 300,535~ETHW tokens in what we term hard-fork-arbitrage. To the best of our knowledge, we are the first to study and quantify hard-fork-arbitrage. 
    \item Finally, we find that the widespread use of ETH staked through LIDO as collateral posed a grave danger to the entire Ethereum blockchain as staking power could have been gobbled up at a significant discount. 
\end{itemize}

\section{Related Work}

DeFi lending emerged in 2017 and first became popularized during the 2020 DeFi summer. Bartoletti et al.~\cite{bartoletti2021sok} provide a systematization of existing knowledge regarding DeFi lending protocols. Further, they offer a formal framework to model the interactions between users in lending pools. In the first empirical study of lending protocols, Gudgeon et al.~\cite{gudgeon2020defi} study different interest rate rules across DeFi lending protocols. Their work analyzes the historical responses of the markets to their liquidity depths. We, on the other hand, investigate the response of lending protocols to the merge and illustrate their reliance on intervention. 

A recent line of work studies the risks stemming from the increasing complexity of DeFi protocol compositions. Tolmach et al.~\cite{tolmach2021formal} provide a formal analysis of DeFi composability and propose a technique for efficient property verification. A measurement study by Kitzler et al.~\cite{kitzler2021disentangling} empirically analyzes and visualizes DeFi compositions -- demonstrating the intertwined nature of DeFi protocols. Wachter et al.~\cite{wachter2021measuring} measure growing asset composability as a proxy for the interoperability of the DeFi applications. This interoperability poses a systemic risk to the DeFi ecosystem, given its resulting convolution. 

The central position of lending in DeFi and the aforementioned intertwined nature of DeFi can lead to increased sensitivity to shocks in the ecosystem. Chiu et al.~\cite{chiu2022inherent} outline that DeFi lending protocols make cryptocurrency prices more sensitive to fundamental shocks. Their work finds that intervention to provide risk management in DeFi lending protocols may improve efficiency and stability while compromising decentralization. In our work, we analyze how actions and interventions taken by DAOs panned out and affected market participants. 

DeFi lending is generally used to facilitate crypto asset price speculation as opposed to real economy lending, as highlighted in a recent bulletin by Aramonte et al.~\cite{aramonte2022defi}. As the authors and others~\cite{xu2022banks,darlin2022debt} point out, the borrowed funds from lending can be reused as collateral to take out additional loans leading to increased leverage. Such leverage spirals, which are in part made possible by DeFi composability,  exasperate the vulnerability of lending protocols towards external events. In contrast to these works, we study the response to and resilience of lending protocols toward external events of lending protocols.  

An active line of research documents and measures multiple attacks and arbitrage opportunities exploiting the design of lending protocols~\cite{eskandari2021sok,qin2021quantifying,yaish2022blockchain}. A particular focus is placed on liquidations. While Perez et al.~\cite{perez2021liquidations} study the efficiency of liquidations, Qin et al.~\cite{qin2021empirical} study optimal liquidation strategies. Our work studies and quantifies a novel and previously unstudied form of arbitrage on lending protocols, which we term hard-fork-arbitrage.

\section{Background}

In the following, we discuss the specifics regarding the merge on the Ethereum blockchain (cf. Section~\ref{sec:merge}), the PoW hard fork (cf. Section~\ref{sec:ethw}), as well as the mechanisms of lending protocols (cf. Section~\ref{sec:lendingprotocol}) and DAOs (cf. Section~\ref{sec:dao}).

\subsection{The Merge}\label{sec:merge}
On 15 September 2022, the merge~\cite{2023themerge} was executed on the Ethereum mainnet. The merge marked the end of energy-intensive PoW and the start of energy-efficient PoS for the Ethereum blockchain. 

PoW, originally proposed by Nakamoto~\cite{nakamoto2008bitcoin}, is the most established consensus for blockchains. Miners must solve a computationally intensive puzzle, and the winner of the puzzle updates the blockchain by appending the newest block. Due to the energy-intensive nature of PoW, there have long been calls to reduce the energy consumption of blockchains. 

The most established blockchain consensus alternative to PoW is the energy-efficient PoS, which was adopted by Ethereum during the merge. Stakers, the miner counterpart for PoS blockchains, offer their funds as collateral for the chance to be selected as a block's validator. For every block, a staker is selected as the block's validator. The chance of being selected as a validator in each round is proportional to their locked-up funds. 

\subsection{EthereumPoW}\label{sec:ethw}

Unsurprisingly, there was uproar from the Ethereum mining community before the merge. The transition rid the miners of their revenue stream and thereby forced them to scrap their hardware or move to other PoW chains. Thus, there were multiple efforts to fork the Ethereum blockchain and create a spinoff, PoW version~\cite{2023EthereumPoW,2023EthereumFair}. Chandler Guo,  a prominent cryptocurrency miner, led the most notable effort to the Ethereum blockchain. The resulting chain is known as EthereumPoW (ETHW)~\cite{2023EthereumPoW}.

Thus, ahead of the merge, a chain split into ETH and ETHW was anticipated. Anyone holding ETH on the original chain right before the merge would automatically receive an equal amount of ETHW tokens after the fork. Speculations surrounding the upcoming chain split led to a trading start of ETHW ahead of the merge. ETHW started trading on 9 August 2022 around 97~US\$ and was trading around 45~US\$ during the merge. 

There have been more than twelve Ethereum hard-forks in the past~\cite{2023EthereumFork}, the most prominent blockchain that resulted from earlier forks being Ethereum Classic (ETC)~\cite{2023ethereumclassic}. This specific fork was anticipated to be particularly challenging for EthereumPoW. Both the rise of DeFi and the prevalence of asset-backed tokens\footnote{Asset-backed tokens are tokens that derive their value from underlying assets that are not necessarily on the same blockchain. E.g., the stablecoin USDC is a token emitted by an organization that promises that for each USDC they hold one US\$.} on the Ethereum chain complicated matters. For example, the organizations behind USDT and USDC, the biggest stablecoins in terms of market capitalization~\cite{2023coinmarketcap}, announced that they would support the transition to PoS~\cite{2022usdtmerge,2022usdcmerge}. As they are both asset-backed, users would hold equal amounts of their USDC/USDT tokens on both chains right after the merge, but the tokens on the ETHW chain would be worthless as they are no longer backed. Note that not only stablecoins but, in fact, the vast majority of ERC20 tokens are asset-backed and were therefore expected to become worthless on the ETHW chain. 

\subsection{Lending Protocols}\label{sec:lendingprotocol}
Lending protocols are among the most successful DeFi applications. They facilitate trustless and decentralized cryptocurrency loans. In traditional finance, one generally receives loans from financial institutions such as banks. As security, banks require collateral for the loan, for example, the house in the case of a mortgage. Additionally, the conditions of the loan are negotiated and vary across loans. DeFi, on the other hand, automates lending by relying on smart contracts. Lending protocols allow anyone to become a lender. By locking their cryptocurrency assets in the protocol's smart contract users become \textit{liquidity providers}. In exchange for providing capital, they earn interest on their assets. Thus, liquidity providers are the lynchpin of lending protocols as they provide the capital.

Interest payments are made by borrowers who take out loans against their locked cryptocurrency collateral. More specifically, borrowers can take out loans without prior clearance by depositing collateral, as long as the collateral is greater in value, i.e., the loans are over-collateralized. Over-collateralization, thus, is the key behind the trustless nature of lending protocols, as it protects lenders against downside price movements of the borrowers' collateral. Once a loan is no longer sufficiently over-collateralized the borrower is incentivized to adjust the position. Generally, loans are insufficiently collateralized if the value of the collateral does not exceed the debt value by more than 20\%. This margin does vary depending on the protocol's risk assessment of the collateral. In case the borrower does not react, the position will likely be closed by \emph{liquidators} at a cost for the borrower.

Note that loans on lending protocols are generally for an indefinite time, as interest payments are made periodically. Additionally, the interest payments are generally variable and typically dependent on the asset borrowed as well as the utilization. The utilization at time $t$ of an asset is given as 
$U_t = {L_t}/{D_t}$,
where $L_t$ denotes the total outstanding loans and $D_t$ denotes the deposits. We will go through the specifics for AAVE and Compound in the following. AAVE V2 and Compound V2 were the largest and newest markets for ETH\footnote{Users technically borrow wrapped ETH (WETH), an ERC20 compatible version of ETH, on the two protocols. As WETH is simply a wrapped version of ETH that has virtually the same value as ETH, we refer to WETH as ETH throughout.} borrowing around the time of the merge. Together they currently account for more than 85\% of the volume locked in lending protocols~\cite{2023lending}.

Borrowers on AAVE can choose between stable interest payments and variable interest rate payments. Both are charged periodically, at every time step, and depend on the asset's utilization. Note that the interest rate is charged periodically by simply adjusting the balance of the debt tokens held by the borrowers (debt is compounded). The interest rate a borrower is charged at time $t$ is given as follows
\begin{equation*}
r_t = \begin{cases}
        r_0 + \dfrac{U_t}{U_{\text{optimal}}} r_{\text{slope}_1} & \text{if }U_t \leq U_{\text{optimal}},  \\
        r_0 + r_{\text{slope}_1} +\dfrac{U_t-U_{\text{optimal}}}{1-U_{\text{optimal}}} r_{\text{slope}_2} & \text{if }U_t > U_{\text{optimal}}, \\
       \end{cases}
\end{equation*}
where $U_t$ is the current utilization of the asset, and $r_0$, $r_{\text{slope}_1}$, $r_{\text{slope}_2}$, $U_{\text{optimal}}$ are configuration parameters. $U_{\text{optimal}}$ is the target utilization of the protocol, once the utilization rises beyond $U_{\text{optimal}}$ borrowing rates rise sharply. Note that both the stable and variable interest rates are computed as indicated above, but the configuration parameters for the same asset differ. The configuration parameters for ETH were set as indicated in Table~\ref{tab:aavepara} during the merge. Further, we draw both the stable and variable interest rates as a function of the utilization in Appendix~\ref{app:interestrate} (cf. Figure~\ref{fig:ratesutilization}). Once the utilization surpasses $U_{\text{optimal}}$, the interest rises at a significantly higher rate, i.e., there is a \emph{kink} in the interest rate curve at $U_{\text{optimal}}$. A loan that is taken out with a variable interest rate is charged periodically according to the current variable interest rate. On the other hand, a loan that is taken out with a stable interest rate at time $t$, $r^s_t$, continues to be charged this rate. Note that the stable rate is not guaranteed to remain stable for an indefinite time period. Instead, it can be adjusted if the loan's stable rate is lower than the current supply rate received by lenders~\cite{Frangella2022Aave}. 

\begin{table}[ht]
    \centering
    \begin{subtable}[t]{0.605\linewidth}
        \centering
        \begin{tabular}[t]{@{}llllll@{}}
        	\toprule
        	\textbf{}  & \textbf{$U_{\text{optimal}}$} & \textbf{$r_0$} &\textbf{$r_{\text{slope}_1}$} &\textbf{$r_{\text{slope}_2}$} & \textbf{$R$}\\
        	\midrule
            \textbf{stable rate} & 70\% &3\% &4\% &100\%&10\%\\
            \textbf{variable rate} & 70\% &  0\%& 3\% &100\%&10\% \\
            \bottomrule           
        \end{tabular} 
        \vspace{0.05cm}
        \caption{AAVE} \label{tab:aavepara}
    \end{subtable}
    \hfill
    \begin{subtable}[t]{0.378\linewidth} 
        \centering
        \begin{tabular}[t]{@{}p{1.99cm}  p{0.54cm}  p{0.94cm}  p{0.57cm} @{}}
        	\toprule
        	\textbf{}  &  \textbf{$r_0$} &\textbf{$r_{\text{slope}_1}$} & \textbf{$R$}\\
        	\midrule
            \textbf{variable rate} & 2\% &10\% &20\%\\
            \bottomrule           
        \end{tabular} 
        \vspace{0.28cm}
        \caption{Compound V2} \label{tab:compoundpara}
    \end{subtable}    
    \caption{Parameters for ETH on AAVE V2 and Compound ahead of the merge. Note that the Compound parameters were adjusted on 10 September 2022 (cf. Section~\ref{sec:compound}).}\label{tab:para}\vspace{-10pt}    
\end{table}

Lenders, on the other hand, deposit their assets and in return receive continuous interest rate payments. Precisely, the supply rate, that is the rate they receive, at time $t$ is given as
\begin{equation*}
s_t = U_t (D^s_t\tilde{r}^s_t + D^v_t r^v_t)(1-R),
\end{equation*}
where $D^s_t$ is the share of stable loans, ${r}^s_t $ is the average stable interest rate, $D^v_t$ is the share of variable loans, and $r^v_t$ the variable interest rate. Further, $R$ is the reserve factor, which signifies the minimum proportion of borrow rate payments that flow into the protocol's treasury. Thus, the supply rate is always lower than the borrowing rate, especially when utilization is low. The difference between the two rates is the revenue source of the protocol. We note that lenders can withdraw their assets at all times, as long as the utilization allows for it, i.e., there are sufficient funds that are currently not being borrowed.

As opposed to AAVE, Compound only offers variable interest rate loans. Furthermore, while AAVE indicates annualized rates and then charges for the time the money was borrowed, Compound charges a per-block rate. However, the Compound smart contract is configured with yearly rates and assumes 2,102,400 blocks per year~\cite{2023CompoundEtherscan}. Throughout this work, whenever we display annualized rates for Compound, for better comparability with AAVE, we will assume 6,245 blocks per day (2,279,425 blocks per year), the average number of daily blocks ahead of the merge. 

For ETH, the interest rate at time $t$ is given as follows
\begin{equation*}
r_t = r_0+ U_t \cdot r_{\text{slope}},
\end{equation*}
where $U_t$ is again the utilization, and $r_0$, $r_{\text{slope}}$ are configuration parameters\footnote{Note that ahead of the merge, ETH adopted Compound's standard interest rate model, i.e., the interest rate increases linearly with the utilization and does not exhibit a change in slope. This, however, was adapted in anticipation of the merge (cf. Section~\ref{sec:mergeant}).}. We provide the parameters ahead of the merge in Table~\ref{tab:compoundpara}. Notice that while the interest rate on Compound is higher than on AAVE for low utilization, it is significantly lower for high utilization. Similarly to AAVE, lenders on Compound deposit their assets and receive interest payments continuously. The supply rate is given as
\begin{equation*}
s_t = r_t \cdot U_t (1-R),
\end{equation*}
where $r_t$ is the borrowing interest rate and the $R$ is again a reserve factor.

On both AAVE and Compound, loans that are close to no longer being sufficiently over-collateralized become available for liquidation. To be precise, if the \emph{health factor} of a position drops below 1, a position can be liquidated. A position's health factor is given as 
\begin{equation*}
    H = \frac{\sum\limits _{i\in A} ( C_i \cdot l_i) }{\sum\limits _{i\in A} D_i},
\end{equation*}
where $A$ is the set of available assets on the platform. $C_i$ is the position's collateral in asset $i$ and $D_i$ is the position's debt in asset $i$. Finally, $l_i$ is the liquidation threshold for asset $i$, which is a configuration parameter. A position with a liquidation threshold of 75\% is considered under-collateralized if the value of the debt rises above 75\% in comparison to the collateral. Once a position becomes available for liquidation, its collateral is auctioned off at a discount if the liquidator repays the debt in return.

\subsection{DAO}\label{sec:dao}
Many DeFi protocols, including AAVE and Compound, are governed by a DAO. A DAO is generally composed of the protocol's token holders, who come together to make decisions regarding the protocol according to specified rules that are enforced by the smart contract. Generally, DAOs can make changes to the protocol itself and make decisions regarding the protocol's funds. 

The AAVE DAO is composed of AAVE (AAVE's native token) holders, while the Compound DAO is composed of COMP (Compound's native token) holders. Both DAOs have the power to change the lending protocol's risk parameters in order to be able to respond to changing risks regarding the market's assets. The community generally first discusses proposed changes and then decides by voting. Depending on the outcome of the vote, the changes will automatically be adopted by the protocol's smart contracts. 

Finally, despite their name, DAOs are only as decentralized as the distribution of governance tokens among the actively participating users. However, an analysis of the voting power of various DAOs has shown that the voting power is generally, in effect, very concentrated~\cite{fritsch2022analyzing}.

\section{Data}

We concentrate the data analysis between 9 August, the day the price for ETHW became available, and 15 October 2022, a month after the merge. With this time frame, we can observe behaviors on lending protocols and their implications. Note that we occasionally include data for shorter or longer time periods to understand general trends better or to zoom in on details. In the following, we provide a concise description of our data collection.

\subsection{Ethereum}
To collect data from the Ethereum blockchain, we run an Erigon~\cite{2022erigon} Ethereum archive node, i.e., a node that builds an archive of historical state. In particular, we collect data from AAVE~\cite{2021aave} and Compound~\cite{2021compound}, the two biggest lending protocols on the Ethereum blockchain that have an ETH borrowing market. To obtain the relevant data from the lending protocols, we filter for event logs emitted by the two regarding the relevant underlying assets. We also query the historical state of the lending markets by calling the implemented functions daily through the web3.eth API~\cite{2023lweb3py}.

Further, we follow the ETH debt borrowed on the two protocols to identify whether the debt was transferred to cryptocurrency exchanges and, thereby, likely sold. We filter through the transaction traces stored on our Erigon archive node. We utilize the Etherscan (Ethereum block explorer) Label Word Cloud~\cite{2023labelwordcloud} to obtain wallet labels and later be able to identify transfers to exchanges.

\subsection{EthereumPoW}

We run a full geth~\cite{2023ETHW} EthereumPoW node to collect EthereumPoW blockchain data. To the best of our knowledge, geth is the only node implementation specifically for EthereumPoW.  

Further, to follow the ETH debt borrowed on the two protocols after the merge, we filter through the transactions stored on our EthereumPoW node. Note that geth does not implement a trace filter. Thus, we filter the transactions and identify ETHW transfers done through a regular transaction, i.e., a transfer from one account to another. We might miss additional transfers in transactions that execute a contract but still obtain a lower bound for the ETHW transferred. We utilize the Etherscan Label Word Cloud~\cite{2023labelwordcloud} and OKLINK~\cite{2023oklink} (EthereumPoW block explorer) to obtain wallet labels. Note that the addresses owned by exchanges before the merge still belong to those exchanges on the EthereumPoW blockchain.

\subsection{Price Data}
We gather hourly price data for the relevant cryptocurrencies from Yahoo Finance~\cite{2023yahoofinance} by interacting with their Python API~\cite{2023yfinance}. We use Yahoo Finance price data instead of the Chainlink price oracle, as Yahoo Finance tracked the ETHW price. However, when discussing a position's health, we will utilize prices from the respective Chainlink price oracle~\cite{2023Chainlink}.

\section{Merge Anticipation}\label{sec:mergeant}

\begin{figure}[t]
\centering
\includegraphics[scale=1]{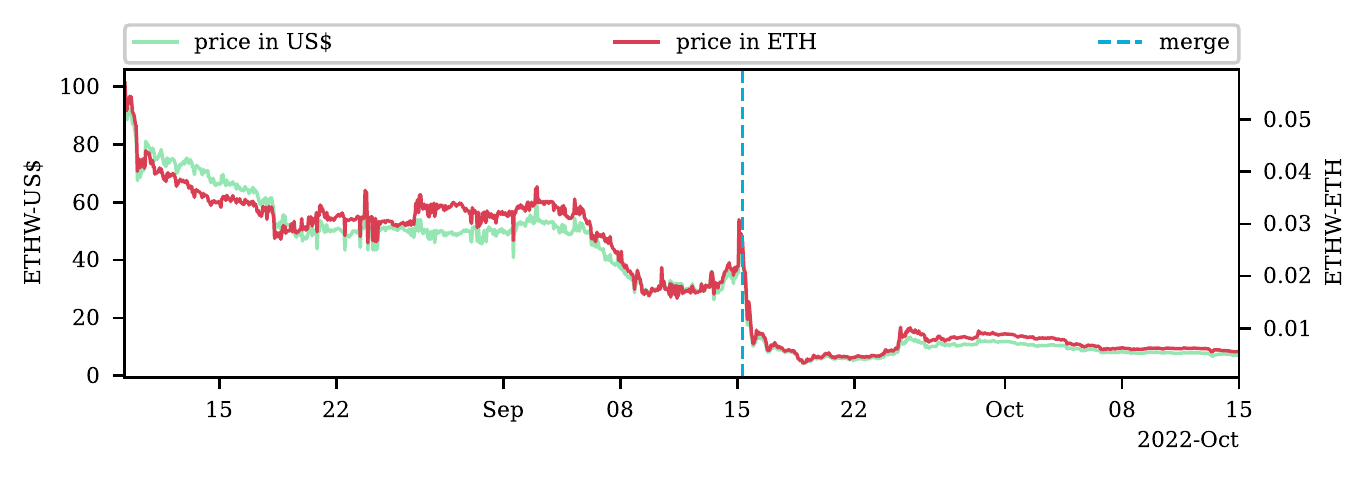}\vspace{-2pt}
\caption{ETHW price in anticipation of and after the merge. The execution of the merge is marked by the dashed blue line. Notice the sharp price increase right before the merge and the dramatic price drop by more than 75\% right after the merge. The line plots are quite similar as the ETH-USD price was significantly less volatile around the merge.}\label{fig:ETHW}\vspace{-2pt}
\end{figure}

We commence the analysis by considering the ETHW price leading up to and post-merge (cf. Figure~\ref{fig:ETHW}). Notice that ETHW's price measured in terms of US\$ (in green) and ETH (in red) moves very similarly. Thus, we infer that the price movements of ETHW were considerably more pronounced than those of ETH. We also see that the price of ETHW generally falls in relation to that of ETH, with one notable exception: right before the merge, ETHW's price measured in ETH spiked. Further, we notice that the value of ETHW was around 3\% that of ETH at the time the merge was executed. Thus, everyone that held ETH right before the merge -- irrespective of whether the ETH was borrowed or not -- received ETHW tokens worth 3\% of their holdings. In particular, one only needed to hold the ETH tokens in one's wallet for a single block, the last block prior to the merge, in order to receive an additional 3\% in value. Thus, in anticipation of this event, lending protocols feared that users would take out ETH loans right before the merge in order to profit from the fork. Note that liquidity providers, with their ETH locked, did not profit from the hard fork and, thus, had the incentive to pull out their funds, further driving up utilization. As excessive borrowing activities can cause major distress to lending protocols, they intervened in order to disincentive/disallow such behavior.
\subsection{Compound}\label{sec:compound}

On 16 August 2022, one month before the merge, the Compound community started discussing the consequences of the merge for their protocol. They devised plans to get ahead of the anticipated event~\cite{monet2022Proposal}. The Compound community raised concerns regarding the liquidity risk for ETH ahead of the merge. In particular, they feared that DeFi users would withdraw ETH from Compound and/or borrow any available ETH. To address this, it was suggested to update the risk parameter and cap the amount of ETH that can be borrowed. After two weeks of discussion, a vote was held by the COMP token holders on 8 September 2022~\cite{ceth2022MonetSupply}. The community voted in favor of the changes, and the alterations were executed two days later, on 10 September 2022. Thus, the process took a total of three weeks. 

As part of the change, Compound switched the interest rate model for ETH to what they call the jump interest rate model. In this model the interest rate at time $t$ contains a kink and is given by 
\begin{equation*}
r_t = \begin{cases}
        r_0+ U_t \cdot r_{\text{slope}_1}& \text{if }U_t \leq U_{\text{optimal}},  \\
        r_0+ U_t\cdot r_{\text{slope}_1} +(U_t-U_{\text{optimal}})\cdot r_{\text{slope}_2} & \text{if }U_t > U_{\text{optimal}}. \\
       \end{cases}
\end{equation*}
Here, $U_t$ is again the asset's current utilization, and $r_0$, $r_{\text{slope}_1}$, $r_{\text{slope}_2}$, $U_{\text{optimal}}$ are configuration parameters (cf. Table~\ref{tab:paraafter}). 

\begin{table}[h]
    \centering
    
    \begin{tabular}{@{}r  l llll@{}}
        \toprule
        \textbf{}  & \textbf{$U_{\text{optimal}}$} & \textbf{$r_0$} &\textbf{$r_{\text{slope}_1}$} &\textbf{$r_{\text{slope}_2}$} & \textbf{$R$}\\
        \midrule
        \textbf{variable rate} & 80\% &2\% &20\% &4910\%&20\%\\
        \bottomrule           
    \end{tabular} 
        \vspace{0.05cm}
    \caption{Parameters for ETH Compound after the adoption of Proposal 122~\cite{monet2022Proposal}.}
    \label{tab:paraafter}\vspace{-8pt}
\end{table}

\begin{figure}[t]
\centering
\begin{subfigure}[t]{1\linewidth}
\includegraphics[scale=1,right]{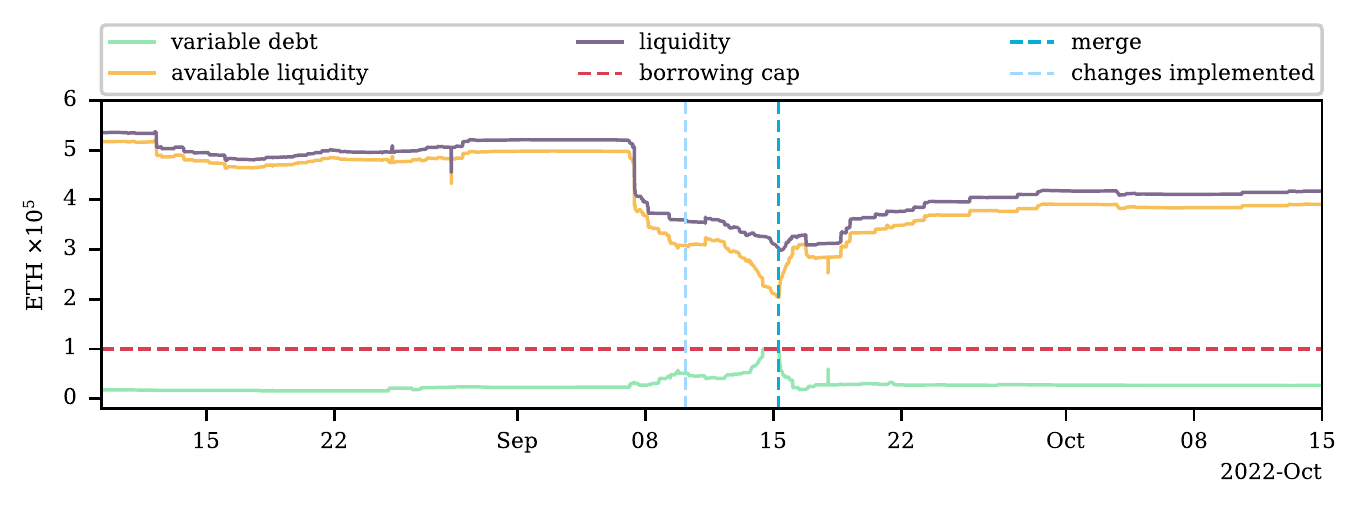}\vspace{-2pt}
\caption{Amount of ETH variable debt, available liquidity, and liquidity on Compound around the merge. The available liquidity is the difference between liquidity and debt. Ahead of the merge, indicated in the plot, Compound imposed a borrowing cap of 100'000 ETH and adjusted the interest rate curve. We indicate the time of the aforementioned changes and show the imposed borrowing cap. Notice that the borrowing cap was reached ahead of the merge. Additionally, liquidity drops ahead of the merge.} \vspace{-2pt}\label{fig:WETH_liquidity_compound}
\end{subfigure}\hfill
\begin{subfigure}[t]{1\linewidth}
\includegraphics[scale=1,right]{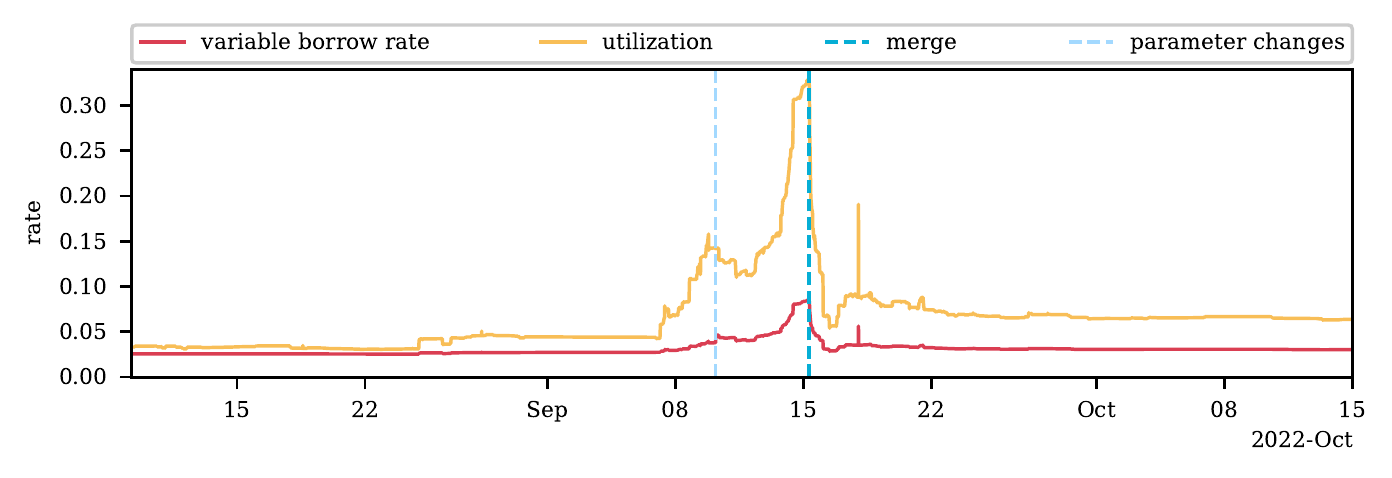}\vspace{-2pt}
\caption{ETH borrow rate and utilization ahead of the merge. Notice the sharp increase in utilization and borrow rate ahead of the merge and the subsequent sharp drop. Yet, the borrow rate and utilization did not reach dramatic levels and stayed below 0.1 and 0.35 respectively.}\label{fig:WETH_rates_compound}\vspace{-2pt}
\end{subfigure}   
\caption{Compound ETH market around the time of the merge. Figure~\ref{fig:WETH_liquidity_compound} shows the market's debt and liquidity, while Figure~\ref{fig:WETH_rates_compound} the borrowing interest rate and utilization.}\label{fig:WETH_compound}\vspace{-8pt}
\end{figure}

To better understand the effect of the imposed changes and the Compound ETH market ahead of the merge in general, we plot the evolution of the ETH debt and liquidity in Figure~\ref{fig:WETH_liquidity_compound}. Further, we show the borrowing rate and utilization over time in Figure~\ref{fig:WETH_rates_compound}. In both plots, we can clearly see the effect of the merge on Compound's ETH lending market. 

Notice that, as expected, two effects play out simultaneously ahead of the merge. For one, we observe an increase in debt, i.e., users appear to borrow ETH in order to be able to exploit this hard-fork-arbitrage opportunity. By borrowing ETH, they can increase the amount of ETH in their wallet and thereby increase the amount of ETHW they will receive once the chains forked. In fact, we looked at all wallets that increased their ETH debt by more than 1,000 ETH between 9 August 2022 and the merge. There were 18 addresses with a debt increase exceeding 1,000 ETH in total, and we could directly trace 50\% of the borrowed funds to cryptocurrency exchanges (cf. Table~\ref{tab:compoundborrowuser}).

When checking whether the addresses transferred the borrowed funds to cryptocurrency exchanges, e.g., Binance, Coinbase, FTX, etc., we monitor ETH(W) transfers from the wallets after they borrowed ETH from Compound. In particular, we check whether any funds were transferred to deposit addresses of exchanges. Exchanges typically have users transfer their assets to deposit addresses, these are created for each user, and then the exchanges forward these funds to their main addresses~\cite{victor2020address}. We search for transfers to exchanges by identifying the transfer of funds from the borrower's address to a deposit address that is then transferred to a known address of an exchange, in transaction data and their trace data on the Ethereum blockchain and EthereumPoW blockchain. We remark that we only filter for these direct transfers to exchanges and thus might miss some additional transfers, where the borrowing address first transferred the funds to another address they control. Further, some of the biggest borrowers ahead of the merge were smart contracts (identified in italics in Table~\ref{tab:compoundborrowuser}) as opposed to externally owned wallets. It is unlikely that the borrowed funds are transferred directly from these smart contracts to exchanges, and we were also never able to identify such a transfer. However, while we might miss some additional transfers to exchanges, only filtering for direct transfers to exchanges allows us to confidently say that the funds were indeed transferred to exchanges by the borrowers and lets us avoid over-counting. Thus, providing us with a lower bound for the total amount transferred to exchanges. 

\begin{table*}
    \centering
    \adjustbox{max width=\textwidth}{

    \begin{tabular}{@{}l  r  l c c c c @{}}
        \toprule
        \textbf{wallet address}  & \textbf{volume [ETH]} & \textbf{exchanges}& \textbf{before}& \textbf{after} & \textbf{>50\%} & \textbf{>99\%}\\
        \midrule
            \texttt{0x712d0f306956a6a4b4f9319ad9b9de48c5345996} &  15,000.00 &          FTX, OKX, MXC, Bybit &       \cmark &      \cmark &   \cmark &   \cmark \\
            \texttt{0xa9f00c00ea5fd167da64917267e60f9d9430b321} &   9,640.00 &                           FTX &       \cmark &      \xmark &   \cmark &   \xmark \\
            \texttt{\textit{0xe40eea78752e969022c3dd18ae68713fd003e1c5}} &   7,771.00 &     \\
            \texttt{0xb0449ec1a8a60f95322617d6ed52e1ba1a7beb49} &   7,000.05 &                           FTX &       \xmark &      \cmark &   \cmark &   \cmark \\
            \texttt{\textit{0x8888882f8f843896699869179fb6e4f7e3b58888}} &   6,611.92 &   \\
            \texttt{0x66b870ddf78c975af5cd8edc6de25eca81791de1} &   5,499.71 & Binance, FTX, OKX, Bybit, MXC &       \cmark &      \cmark &   \cmark &   \cmark \\
            \texttt{0xee8e0fcc8bff03ec5f100d02cb7b3196d78863a7} &   4,499.92 &             FTX, MXC, Binance &       \cmark &      \cmark &   \cmark &   \cmark \\
            \texttt{0x6a704a0e46dcc67a6316644372e261e8fb6f658c} &   3,000.00 &                               &       \xmark &      \xmark &   \xmark &   \xmark \\
            \texttt{0xcfc50541c3deaf725ce738ef87ace2ad778ba0c5} &   2,498.00 &                      Coinbase &       \cmark &      \xmark &   \xmark &   \xmark \\
            \texttt{0x9681319f4e60dd165ca2432f30d91bb4dcfdfaa2} &   2,000.00 &                  FTX, Binance &       \cmark &      \cmark &   \cmark &   \cmark \\
            \texttt{0x5add1cec842699d7d0eaea77632f92cf3f3ff8cf} &   1,665.05 &                           MXC &       \xmark &      \cmark &   \xmark &   \xmark \\
            \texttt{0x42283fa21d5642c1744c2888f041ddea5d79149c} &   1,650.00 &                               &       \xmark &      \xmark &   \xmark &   \xmark \\
            \texttt{0xde6b2a06407575b98724818445178c1f5fd53361} &   1,550.00 &                           OKX &       \xmark &      \cmark &   \cmark &   \cmark \\
            \texttt{0xb5c4402ff7cbe97785dddc768c4e3a4f033474fb} &   1,501.00 &                      FTX, MXC &       \cmark &      \cmark &   \cmark &   \xmark \\
            \texttt{0xf71b335a1d9449c381d867f4172fc1bb3d2bfb7b} &   1,400.00 &                           FTX &       \xmark &      \cmark &   \cmark &   \cmark \\
            \texttt{0x6d68c0f44e86587aa443ddb12ed9f10920195ada} &   1,300.00 &                    OKX, Bybit &       \xmark &      \cmark &   \cmark &   \cmark \\
            \texttt{0xec97b52fc79f9ec7e951f050c80f65cc087197d3} &   1,100.00 &                               &       \xmark &      \xmark &   \xmark &   \xmark \\
            \texttt{0xe7072cdf38d3a6a4b92929abc302325f7b1ca628} &   1,002.00 &                               &       \xmark &      \xmark &   \xmark &   \xmark\\
        \bottomrule           
    \end{tabular} }
        \vspace{0.05cm}
    \caption{We analyze ETH(W) transfers to cryptocurrency exchanges of all addresses (contracts are in italics) whose net borrowing preceding the merge exceeded 1,000 ETH on Compound. Note that the volume column indicates the debt increase. For each wallet, we display the exchanges to which funds were transferred and whether this occurred before or after the merge. Further, we indicate whether the wallet transferred the equivalent in value, at least 50\%, and/or at least 99\% of that debt to exchanges.}\label{tab:compoundborrowuser}\vspace{-20pt}   
\end{table*}

Table~\ref{tab:compoundborrowuser} notes to which exchange(s) funds were transferred by each borrower. Further, we indicate whether the borrower transferred the funds to exchanges before (on the Ethereum blockchain) and/or after (on the EthereumPoW blockchain) the merge. Any transfer to exchanges before the merge was ETH as opposed to ETHW, but some exchanges announced ahead of time that they would give the users ETHW for the ETH held with them~\cite{binancemerge2}. We further indicate in Table~\ref{tab:compoundborrowuser} whether the funds transferred to exchanges amounted to at least 50\% and/or 99\% of the debt taken on by the address ahead of the merge. Most addresses, especially if we disregard the smart contracts where our method does not identify transfers to exchanges, transferred at least half of their new debt to exchanges. While we cannot determine the exact purpose of these transactions, it is highly likely that they intended to sell ETHW. Interestingly, the wallet with the highest debt increase ahead of the merge belonged to the now infamous Alameda Research: the cryptocurrency trading firm that allegedly traded FTX customer funds and lost them. This led to the bankruptcy of FTX in November 2022~\cite{velaquez2022On}. We also want to highlight that we tracked 49,206 ETH(W) to exchanges from only the borrowers shown in Table~\ref{tab:compoundborrowuser} --  more than 49\% of the ETH debt on Compound (100,000 ETH) ahead of the merge. At the time of the merge, the transferred ETHW was worth more than 2 Mio US\$. Note that while borrowers did have to pay interest, these expenses were far lower than the value of ETHW, as is shown in Appendix~\ref{app:cumulative} (cf. Figure~\ref{fig:cumulative_compound}). 

\begin{figure}[t]
\centering
\begin{subfigure}[t]{\linewidth}
\includegraphics[scale=1,right]{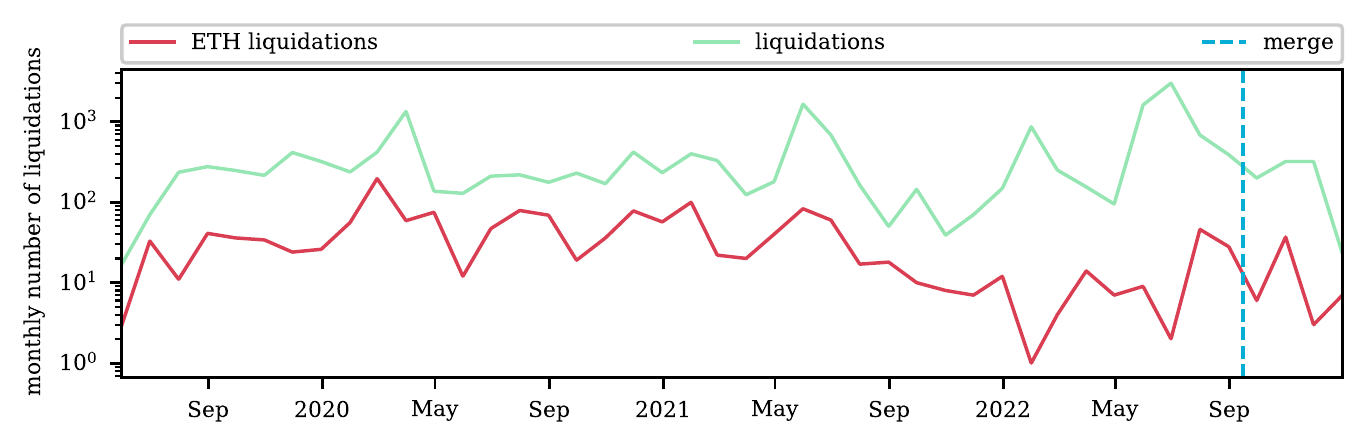}\vspace{-2pt}
\caption{The monthly number of liquidations on Compound. We show in green the total number of liquidations, and in red the number of liquidations where liquidators covered the position's ETH debt.}\vspace{-2pt}
\label{fig:liquidations_compound}
\end{subfigure}\hfill
\begin{subfigure}[t]{1\linewidth}
\includegraphics[scale=1,right]{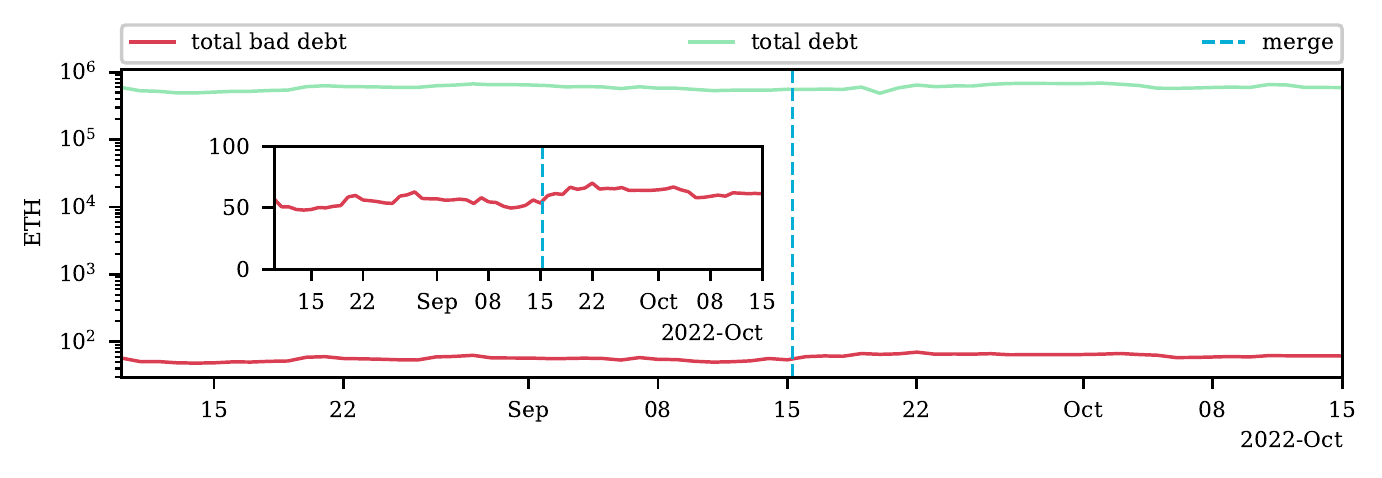}\vspace{-8pt}
\caption{Total debt and bad debt, i.e., debt value exceeds the collateral value, on Compound before and after the merge. The bad debt is a very small proposition of the total debt -- less than 0.01\%. The inset shows a close-up of the bad debt on a linear scale. Notice that the bad debt on Compound does not spike up ahead of the merge.}\vspace{-2pt} \label{fig:bad_debt_compound}
\end{subfigure}   
\caption{Stability of ETH borrowing on Compound. Figure~\ref{fig:liquidations_compound} shows the number of liquidations over time and Figure~\ref{fig:bad_debt_compound} shows the total and bad debt around the time of the merge. An increase in liquidations was avoided on Compound.}\label{fig:liquidations_bad_debt_compound}\vspace{-8pt}
\end{figure}

Besides observing an increase in debt, especially by the 18 addresses discussed previously, we also note a decrease in liquidity (cf. Figure~\ref{fig:WETH_liquidity_compound}). The reasons behind this decrease are likely more complex. For one, lenders might fear a rise in utilization in Compound's ETH market. When utilization levels are very high, lenders can no longer withdraw their funds (liquidity risk). Additionally, lenders might also wish to receive ETHW on the forked chain. However, ahead of the merge, the official ETHW Twitter account recommended for funds be withdrawn from multiple DeFi pools~\cite{EthereumPoW2022EthereumPoW}, including Compound's ETH market, in order to ensure that they would receive ETHW. There was even talk about freezing DeFi contracts on the ETHW fork~\cite{EthereumPoW2022Params}. However, if utilization in Compound's ETH market was high, it is unlikely that it would decrease after the hard fork. Borrowers have no incentives to repay their ETHW debt on the EthereumPoW fork, as their collateral assets there are likely worthless. Thus, users would never be able to withdraw their ETH from Compound on the EthereumPoW chain even if they were not frozen. Despite a noticeable decrease in liquidity, the market's liquidity remained significantly larger than the protocol's debt, which reached the borrowing cap of 100,000 ahead of the merge.

Thus, the intervention by the Compound community ensured that the protocol's utilization remained relatively low, i.e., it never exceeded 35\% (cf. Figure~\ref{fig:WETH_rates_compound}). As a consequence, the borrowing rate also remained relatively low. We further note that Compound did not experience an increase in liquidations of positions ahead of the merge. Figure~\ref{fig:liquidations_compound} shows the total number of monthly liquidations and the share of those liquidations that had ETH debt covered by the liquidators. We presume that it was the intervention by the Compound community that helped prevent mass liquidation as the borrowing rate never exceeded 10\%. 

In addition to averting liquidation, the amount of bad debt on Compound, i.e., positions whose debt value exceeds the collateral value, did not increase significantly in the lead-up to the merge (cf. Figure~\ref{fig:liquidations_compound}). Bad debt can be detrimental to a lending protocol, as the respective loans become irretrievable and present a loss for lenders. While generally, positions become liquidated before the debt value exceeds the collateral, extreme price swings can leave insufficient time to liquidate the positions. Yet, in this case, the stable amount of bad debt and the overall small share of bad debt (less than 0.01\%) relative to the total debt on the protocol indicates that the increased rates did not impact the protocol's health significantly.

\begin{figure}[t]
\centering
\includegraphics[scale=1,right]{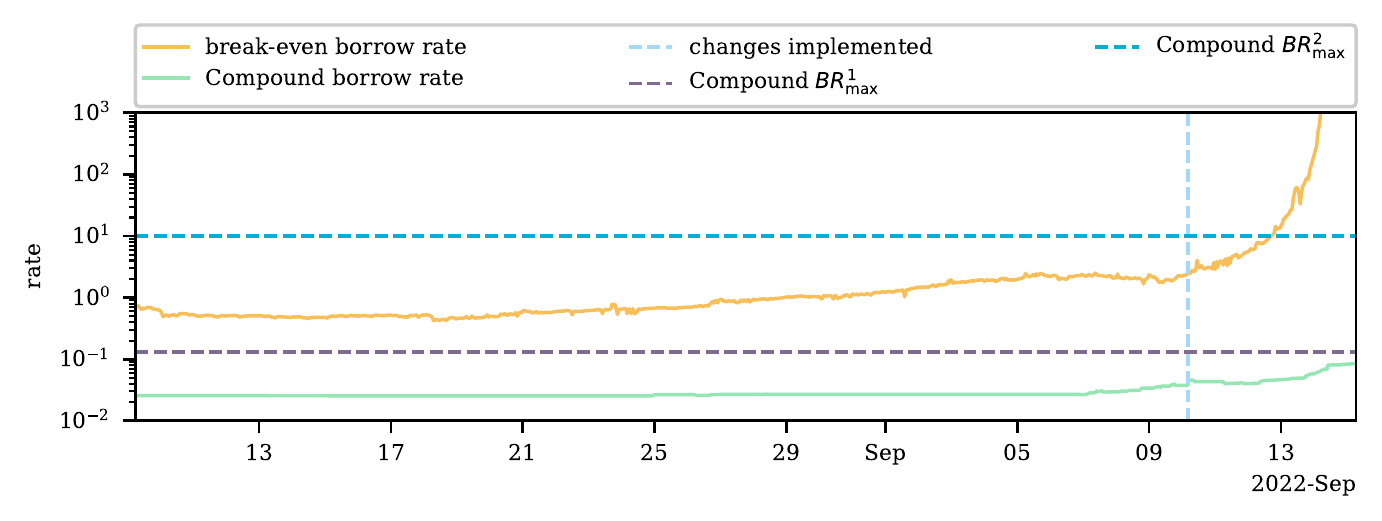}\vspace{-2pt}
\caption{Break-even borrowing rate, i.e., the annualized interest rate users are willing to pay until the merge in return for ETHW, compared to Compound's borrowing rate. We indicate the maximum borrow rate before, $BR_{\max}^1$, and after the implemented changes, $BR_{\max}^2$. Note that Compound's borrowing rate was always at least one order of magnitude smaller than the break-even borrow rate.} \label{fig:breakevenborrowrate_compound}\vspace{-8pt}
\end{figure}

At the same time, the actions of the Compound community ensured that those who did manage to take out loans in time, i.e., before the borrowing cap was reached, could make significant profits. In Figure~\ref{fig:breakevenborrowrate_compound}, we plot the break-even borrowing rate in the lead-up to the merge along with the actual Compound borrowing rate. The break-even borrowing rate at time $t$ indicates the annualized rate a user would be willing to pay for borrowing ETH between time $t$ and the merge, given the relative price between ETHW and ETH at time $t$. Note that we take the futures price at time $t$ and not the price ETHW started trading at post-merge, as a borrower at time $t$ could not know this price but rather had to rely on the price of future contracts. We compute the break-even borrow rate at time $t$ as follows
\begin{equation*}
    \left(1 + \frac{p_{\text{ETHW}}(t)}{p_{\text{ETH}}(t)}\right) ^{\frac{\Delta_{\text{year}}}{\Delta_{\text{merge}}}}-1,
\end{equation*}
where $p_{\text{ETHW}}(t)$ is the price of ETHW in US\$, $p_{\text{ETH}}(t)$ is the price of ETH in US\$, $\Delta_{\text{year}}$ is the number of seconds in a year, and $\Delta_{\text{merge}}$ the number of seconds until the merge. Notice that at all times, the break-even borrow rate exceeded the actual borrow rate by at least one order of magnitude. Further, the break-even borrow rate was, for the most part, smaller than the maximum possible borrow rate on Compound. We show the maximum borrow rate before the imposed changes, $BR_{\max}^1$,  in violet and the maximum borrow rate after the imposed changes, $BR_{\max}^2$, in light blue. Thus, users could infer that with a very high probability the short-term borrowing costs would be significantly lower than the value of the ETHW, they would receive once the chains forked. As a result, users were enticed to take leveraged long positions if they still could, i.e., until the borrowing cap was reached. Importantly, the opposite was true for lenders. They were insufficiently compensated and would have been better off withdrawing the funds and directly profiting from the merge.

\subsection{AAVE}

The AAVE community started discussing the possible repercussions of the merge on 23 August 2022~\cite{Primoz2022arc}. They were also concerned that users would borrow as much ETH as possible to maximize their ETH holdings in anticipation of the fork. Such activity would increase utilization, make liquidations harder, and possibly lead to ETH suppliers withdrawing their ETH from the platform. 

An additional challenge for AAVE, as opposed to Compound, was that they allow \emph{staked ETH} (stETH) -- the token users receive in exchange for staking their ETH with LIDO -- as collateral. LIDO~\cite{2023lido} is a protocol that allows you to easily stake your ETH on the PoS consensus layer -- the Beacon chain. Normally, in order to stake ETH on the Beacon chain, users require 32 ETH, but LIDO is a liquid staking solution that allows its users to stake any amount. Thus, increasing accessibility to ETH staking. Staking rewards received by the ETH staked through LIDO on the Beacon chain are distributed to the users on a daily basis. More precisely, LIDO updates its Beacon chain balance every 24 hours on the Ethereum mainnet. The stETH balances in the wallets automatically update accordingly. Thus, stETH is an interest-earning token. stETH can be bought and sold by users.

Importantly, staking on the Beacon chain was activated more than a year ahead of the merge. Thus, many LIDO users had staked their ETH and received stETH in return ahead of the merge. Some stETH holders decided to utilize their stETH as collateral to take out ETH loans on AAVE, proceeded to stake the ETH they borrowed on LIDO, again received stETH, and continued this process. The staking rewards received by stETH holders historically exceeded the ETH borrowing rates on AAVE, making the aforementioned strategy profitable. The AAVE community feared liquidations of these positions as ETH borrowing rates were anticipated to rise ahead of the merge. 

In their attempt to mitigate such scenarios, the AAVE community took a different route than the Compound community. They decided to pause all ETH lending on the platform ahead of the merge. A vote regarding the proposal was held between 2 and 6 September 2022~\cite{Lei2022Pause}. As the community was in favor of the changes, they were implemented on 7 September 2022 --- a good week ahead of the merge. Thus, from 7 September onward, it was no longer possible to borrow ETH on AAVE.

\begin{figure}[ht]
\centering
\begin{subfigure}[t]{\linewidth}
\includegraphics[scale=1,right]{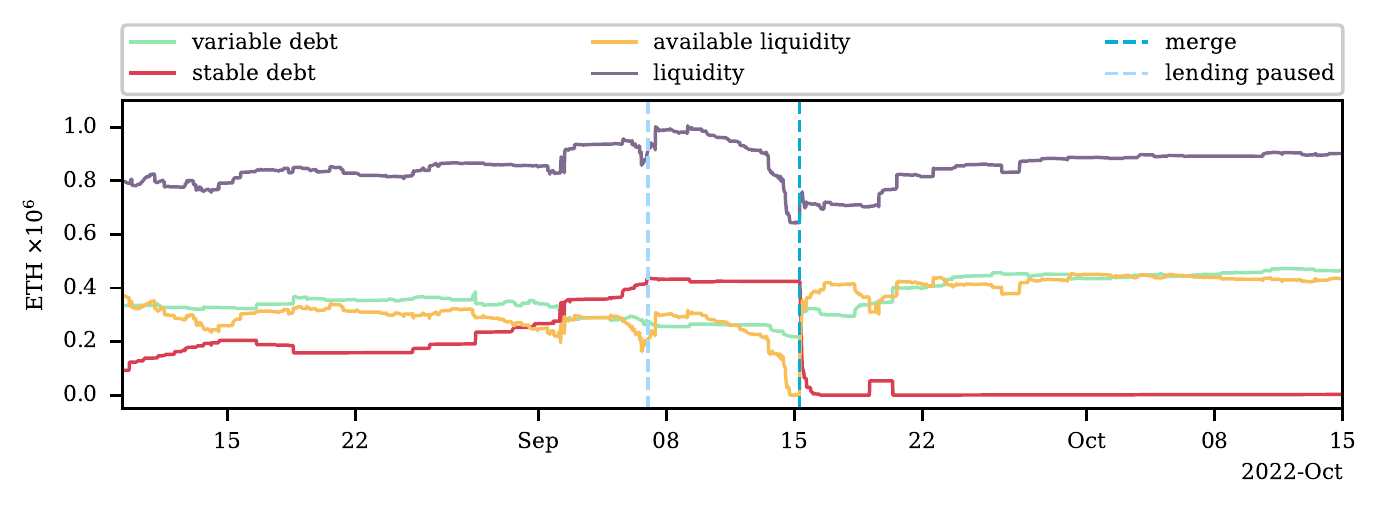}\vspace{-2pt}
\caption{Amount of ETH variable debt, stable debt, available liquidity, and liquidity on AAVE around the merge. The available liquidity is the difference between liquidity and debt. Ahead of the merge AAVE paused ETH borrowing. We indicate the time of the aforementioned changes. Notice that the available liquidity continues to sink even though borrowing was paused as lenders are leaving the market. Furthermore, the slight increase in debt even after lending was paused is due to the fact that interest is accumulated on Aave.} \label{fig:WETH_liquidity_aave}\vspace{-2pt}
\end{subfigure}\hfill
\begin{subfigure}[t]{\linewidth}
\includegraphics[scale=1,right]{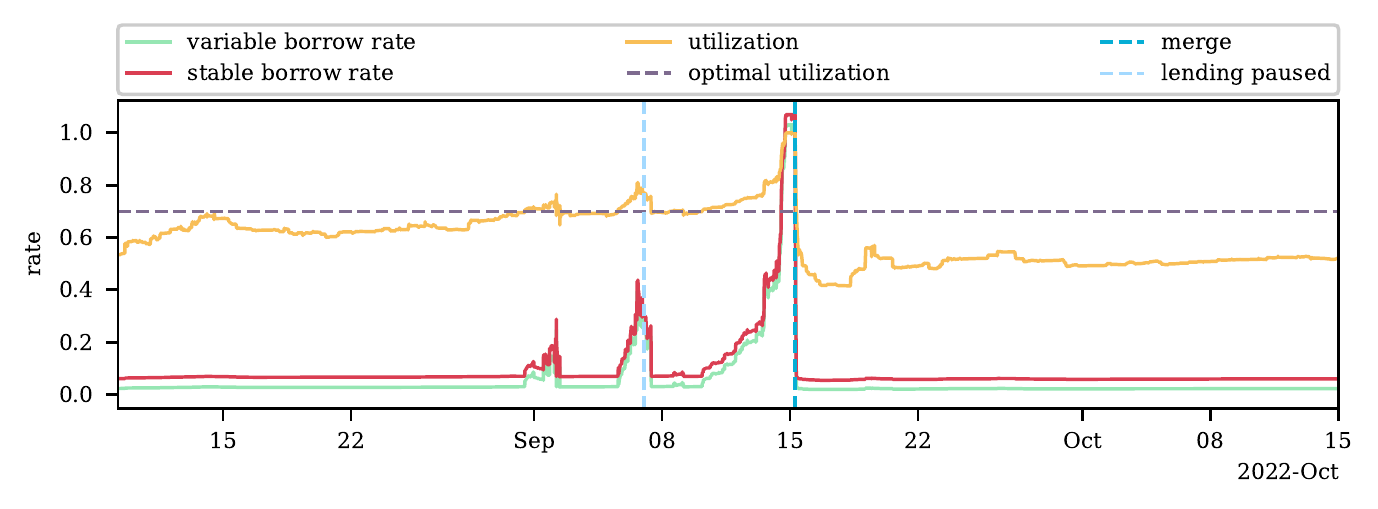}\vspace{-2pt}
\caption{The ETH variable and stable borrow rate, as well as utilization ahead of the merge. Notice the sharp increase in utilization and borrow rate ahead of the merge and the subsequent sharp drop. The utilization of AAVE's ETH market reached 100\% ahead of the merge even though lending was paused.} \label{fig:WETH_rates_aave}\vspace{-2pt}
\end{subfigure}   
\caption{AAVE ETH market around the time of the merge. Figure~\ref{fig:WETH_liquidity_aave} shows the market’s debt and liquidity, while Figure~\ref{fig:WETH_rates_aave} the borrowing interest rate and utilization.}\label{fig:WETH_aavev2}\vspace{-2pt}
\end{figure}

As shown in Figure~\ref{fig:WETH_liquidity_aave} these changes prevented a further increase in borrowing. In particular, the borrowed stable debt volume (red) increased until 7 September but remained basically flat from the implementation of these changes to the merge. Post-merge, the borrowed volume drops dramatically, highlighting that this borrowing activity was predominantly fuelled by speculators betting on the ETHW windfall. Thus, as plotted in Figure~\ref{fig:WETH_rates_aave}, the implemented changes failed to keep borrowing rates in check. Liquidity providers enticed to directly hold ETH to themselves profit from the merge, withdrew their funds from the ETH pool, thus driving up the utilization rate. Additionally, the general uncertainty surrounding lending platforms in the weeks leading up to the merge may have spooked some lenders, further increasing outflows. We note that even with this intervention the utilization reached 100\%, but the time it took to reach this level was likely prolonged. If the utilization had reached this level earlier, total interest payments would have been higher, thus compensating lenders more fairly.

\begin{table*}[ht]
\adjustbox{max width=\textwidth}{
    \centering    
    \begin{tabular}{@{}l  r  l c c c c @{}}
        \toprule
        \textbf{wallet address}  & \textbf{volume [ETH]} & \textbf{exchanges}& \textbf{before}& \textbf{after} & \textbf{>50\%} & \textbf{>99\%}\\
        \midrule
        \texttt{0xd275e5cb559d6dc236a5f8002a5f0b4c8e610701} &  49,998.89 &                      Bitfinex &       \cmark &      \xmark &   \cmark &   \xmark \\
        \texttt{0x54dda22ae140edb605c73073eabb6f4aea2fc237} &  39,999.57 &                  Binance, FTX &       \cmark &      \xmark &   \cmark &   \cmark \\
        \texttt{0xcde35b62c27d70b279cf7d0aa1212ffa9e938cef} &  22,762.39 &                      OKX, FTX &       \cmark &      \cmark &   \cmark &   \cmark \\
        \texttt{0x236f233dbf78341d25fb0f1bd14cb2ba4b8a777c} &  17,500.00 &                               &       \xmark &      \xmark &   \xmark &   \xmark \\
        \texttt{0x68963dc7c28a36fcacb0b39ac2d807b0329b9c69} &  16,022.93 &                           FTX &       \cmark &      \xmark &   \cmark &   \cmark \\
        \texttt{0xf6bf776c06a9946a7beba3bacbdaeb44e90684e1} &  15,000.00 &                           FTX &       \cmark &      \xmark &   \cmark &   \xmark \\
        \texttt{0x68030330e8158be3fa5b3ec3c94bf07e42824b9b} &  14,894.68 &      Binance, Bybit, OKX, FTX &       \cmark &      \cmark &   \cmark &   \cmark \\
        \texttt{\textit{0x5beabefb832db8c0f5a2370b447613c8ebe572eb}}&   8,951.73 &         \\
        \texttt{0x2bde0f6bfc26389fadccee7c1ca14bbf29c45812} &   7,520.00 &                               &       \xmark &      \xmark &   \xmark &   \xmark \\
        \texttt{\textit{0x4256886373b79e4e12c12b6796e99cde90f5f236}} &   7,353.31 &       \\
        \texttt{\textit{0x321bf29f2d5dad204b5e25c31cac4348b6f29f96}} &   7,304.10 &                        \\
        \texttt{0x8d8b9c79196f32161bcb2a9728d274b3b45eb9af} &   7,051.00 &                               &       \xmark &      \xmark &   \xmark &   \xmark \\
        \texttt{\textit{0xe40eea78752e969022c3dd18ae68713fd003e1c5}} &   6,950.00 &                    \\
        \texttt{\textit{0x4c1bda12452146184a8085c890e22fb7933aff2f}} &   6,250.00 &                    \\
        \texttt{0x48608b596888e0b9512be7f3f5f2e05d3c3d5180} &   5,300.00 &                           FTX &       \cmark &      \xmark &   \cmark &   \cmark \\
        \texttt{0xcc8a1601a32b48cebf45224ca6d786c24414a10b} &   4,500.00 &                           FTX &       \cmark &      \xmark &   \xmark &   \xmark \\
        \texttt{\textit{0x2a0fe598e69a4fc882f6f7a954662cf0a0819467}} &   4,460.18 &                  \\
        \texttt{0xca00bf9fa7bee6034565bf5d8e7f95fe52182241} &   4,200.00 &                               &       \xmark &      \xmark &   \xmark &   \xmark \\
        \texttt{0x1bda63dab1743089af8c0c94ed0b75772a9b9858} &   4,000.40 &                       Binance &       \cmark &      \xmark &   \cmark &   \cmark \\
        \texttt{0xadbab4f38ff9dcd71886f43b148bcad4a3081fb9} &   3,998.62 &                           MXC &       \xmark &      \cmark &   \xmark &   \xmark \\
        \texttt{0x916792f7734089470de27297903bed8a4630b26d} &   3,768.00 &                           FTX &       \xmark &      \cmark &   \cmark &   \cmark \\
        \texttt{0xe8b22a88deb45c7848d394fd039b8d811511a9f3} &   3,000.00 &             Binance, OKX, FTX &       \cmark &      \xmark &   \cmark &   \cmark \\
        \texttt{\textit{0x9b1945d5434b2e69eb00e44b9022ad4172922eb5}} &   2,999.00 &                       \\
        \texttt{\textit{0x2662d826a86d602c01affd6974432e43009eb14b}} &   2,729.17 &                           \\
        \texttt{0xb1473f4d2e416310e4715cc7bcbe8074aed24a56} &   2,200.00 &                    Bybit, OKX &       \xmark &      \cmark &   \cmark &   \cmark \\
        \texttt{0xb5c4402ff7cbe97785dddc768c4e3a4f033474fb} &   2,180.00 &                      MXC, FTX &       \cmark &      \cmark &   \cmark &   \xmark \\
        \texttt{0x3da0ca6c78ea283200a0d5b2790aa5de280e43cc} &   2,000.00 &                               &       \xmark &      \xmark &   \xmark &   \xmark \\
        \texttt{0x66b870ddf78c975af5cd8edc6de25eca81791de1} &   1,998.65 & Binance, Bybit, FTX, MXC, OKX &       \cmark &      \cmark &   \cmark&    \cmark \\
        \texttt{0x1778767436111ec0adb10f9ba4f51a329d0e7770} &   1,711.01 &                           FTX &       \cmark &      \xmark &   \cmark &   \cmark\\
        \texttt{0xa1175a219dac539f2291377f77afd786d20e5882} &   1,600.00 &                               &       \xmark &      \xmark &   \xmark &   \xmark \\
        \texttt{0x474e2cb1aac71f66d0aa7adb0cd92c919f842fe4} &   1,599.88 &      Binance, MXC, Bybit, FTX &       \cmark &      \cmark&    \cmark &   \cmark\\
        \texttt{0x7f960b97b12ef8b6828529e961f6646ad764d90b} &   1,500.00 &                           MXC &       \xmark &      \cmark &   \cmark &   \cmark \\
        \texttt{\textit{0x307111465e4cedd89fa28b9768981b8768a3cabe}} &   1,400.00 &                       \\
        \texttt{0x09d0ed8d3ebf0b0b5d2a3d7096546d6d7085b8bb} &   1,364.00 &                           FTX &       \xmark &      \cmark &   \cmark &   \cmark \\
        \texttt{\textit{0x74b8c7680502931c33d9446e26592b8318eb7248}} &   1,110.99 &                       \\
        \texttt{0x3d9663bbd7f238b940ad4244fac58ff54ce870dc} &   1,100.00 &                  Binance, FTX &       \cmark &      \xmark &   \cmark &   \cmark \\
        \texttt{0xecfb36305daa4244281d8249783bddf0918db361} &   1,016.00 &                  Binance, FTX &       \cmark &      \xmark &   \cmark &   \cmark \\
        \texttt{0x7ce450c2974746e3d21b13cb05d253e6fd56f6bd} &   1,000.00 &                           OKX &       \xmark &      \cmark &   \cmark &   \cmark \\
        \texttt{0x05f65845a202aadabce5475b6495f54fb2073b04} &   1,000.00 &                        Peatio &       \cmark &      \xmark &   \cmark &   \xmark \\

        \bottomrule           
    \end{tabular} }
        \vspace{0.05cm}
    \caption{We analyze ETH(W) transfers to cryptocurrency exchanges of all addresses (contracts are in italics) whose net borrowing in the lead-up to the merge exceeded 1,000 ETH on AAVE. Note that the volume column indicates the debt increase. For each wallet we display the exchange(s) to which funds were transferred and whether this occurred before or after the merge. Further, we indicate whether the wallet transferred the equivalent in value, at least 50\%, and/or at least 99\% of that debt to exchanges.}   \label{tab:aaveborrowuser}\vspace{-20pt} 
\end{table*}

Similarly, as done for Compound in Section~\ref{sec:compound}, we again track the funds borrowed by the biggest AAVE borrowers ahead of the merge to see whether the funds were transferred to cryptocurrency exchanges -- indicating that the borrowers wanted to sell the ETHW. In Table~\ref{tab:aaveborrowuser}, we indicate whether and when funds were transferred to exchanges for all 38 ETH borrowers on AAVE that increased their debt by more than 1,000 ETH ahead of the merge. We highlight smart contract borrowers in italics and note again that for these we were not able to track direct transfers to cryptocurrency exchanges. Focusing on the borrowers whose addresses were externally owned wallets, we find that more than 72\% of those transferred at least 50\% of the debt they took on directly to cryptocurrency exchanges, while 58\% transferred at least 99\%. Further, we find that these 29 borrowers moved 39\% (251,329 ETH(W)) of all funds borrowed on AAVE (643,367 ETH(W)) to cryptocurrency exchanges. The transferred ETHW amounted to more than 11 Mio US\$ at the time of the merge. The lenders were essentially deprived of these funds, as they were not fairly compensated for their service as liquidity providers. As we illustrate in Appendix~\ref{app:cumulative} in Figure~\ref{fig:cumulative_aave}, the borrowing costs were far lower than the value of ETHW.

\begin{figure}[ht]
\centering
\begin{subfigure}[t]{1\linewidth}
\includegraphics[scale=1,left]{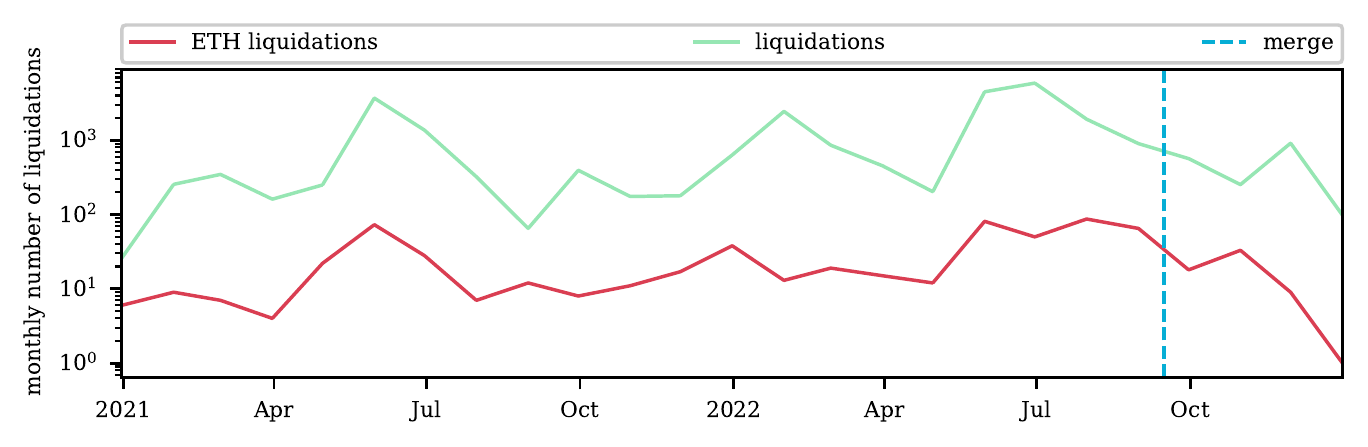}
\caption{The monthly number of liquidations on AAVE. We show in green the total number of liquidations, and in red the number of liquidations where liquidators covered the position's ETH debt.} \label{fig:liquidations_aave}
\end{subfigure}\hfill
\begin{subfigure}[t]{1\linewidth}
\includegraphics[scale=1,left]{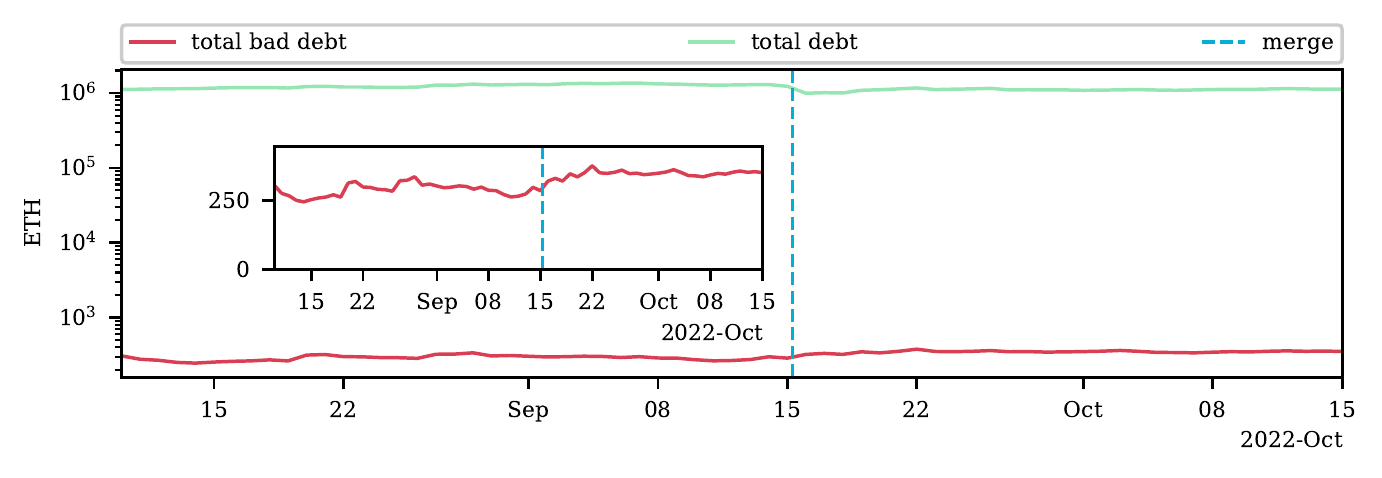}
\caption{Total and bad debt, i.e., debt value exceeds collateral value, on AAVE before and after the merge. The bad debt is a very small proposition of the total debt -- less than 0.04\%. Notice that the bad debt on AAVE does not spike up ahead of the merge.} \label{fig:bad_debt_aave}
\end{subfigure}   
\caption{Stability of ETH borrowing on Aave. Figure~\ref{fig:liquidations_aave} shows the number of liquidations over time and Figure~\ref{fig:bad_debt_aave} shows the total and bad debt around the merge. As with Compound, an increase in liquidations was avoided.}\label{fig:liquidations_bad_debt_aave}\vspace{-2pt}
\end{figure}

While the borrowing rates on AAVE increased significantly before the merge, the worst fears of mass liquidations did not materialize. As shown in Figure~\ref{fig:liquidations_aave}, the number of liquidations occurring on AAVE did not rise significantly in the lead-up to the merge. Similarly, as plotted in Figure~\ref{fig:bad_debt_aave}, no significant increase in the proportion of bad debt can be observed. While the utilization rate did spike, this only persisted for a short time, and as the maximal borrowing rate is capped at 103\% on an annualized basis, the total interest expense for borrowers was manageable.

\begin{figure}[ht]
\centering
\includegraphics[scale=1,right]{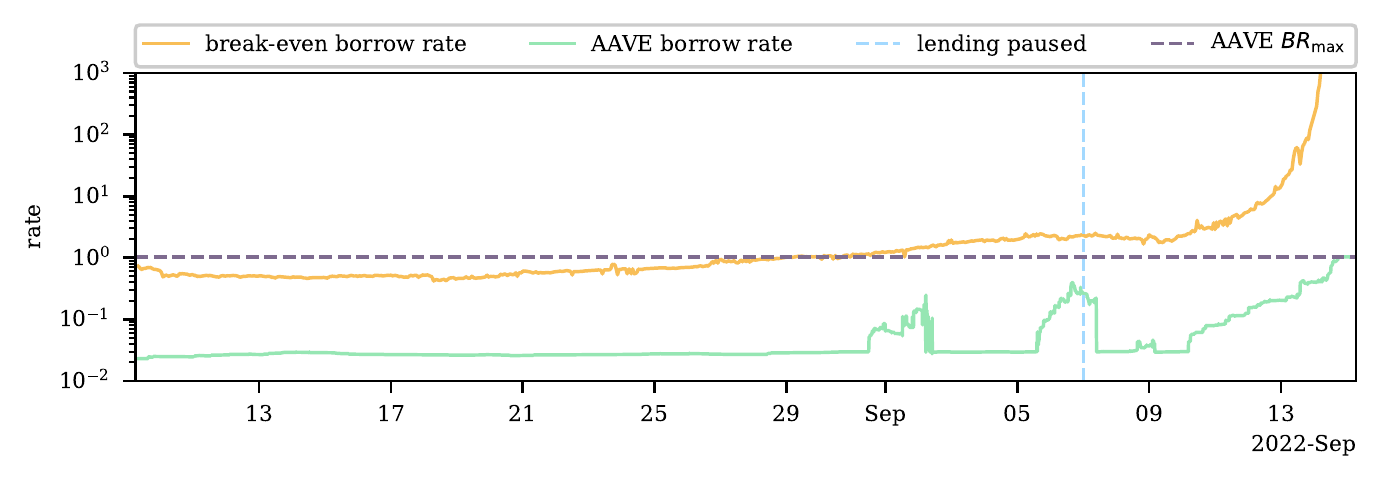}\vspace{-2pt}
\caption{Break-even borrowing rate, i.e., the annualized interest rate users are willing to pay until the merge in return for ETHW, in comparison to variable borrow rate on AAVE. We further indicate the protocol's maximum borrow rate, $BR_{\max}$. Notice the significant discrepancy between the AAVE variable borrow rate and the break-even borrow rate.} \label{fig:breakevenborrowrate_aave}\vspace{-2pt}
\end{figure}

In Figure~\ref{fig:breakevenborrowrate_aave}, we again plot the break-even rate (yellow). As for Compound, the actual borrowing rates were significantly lower than the break-even rates, making leveraged long positions in ETH profitable. Given that ETHW futures traded at about 3\%, the maximal annual borrowing rate of 103\% that AAVE allows was orders of magnitude lower than the break-even rate. Thus, the protocol again inadequately compensated lenders.

\begin{figure}[ht]
\centering
\includegraphics[scale=1,right]{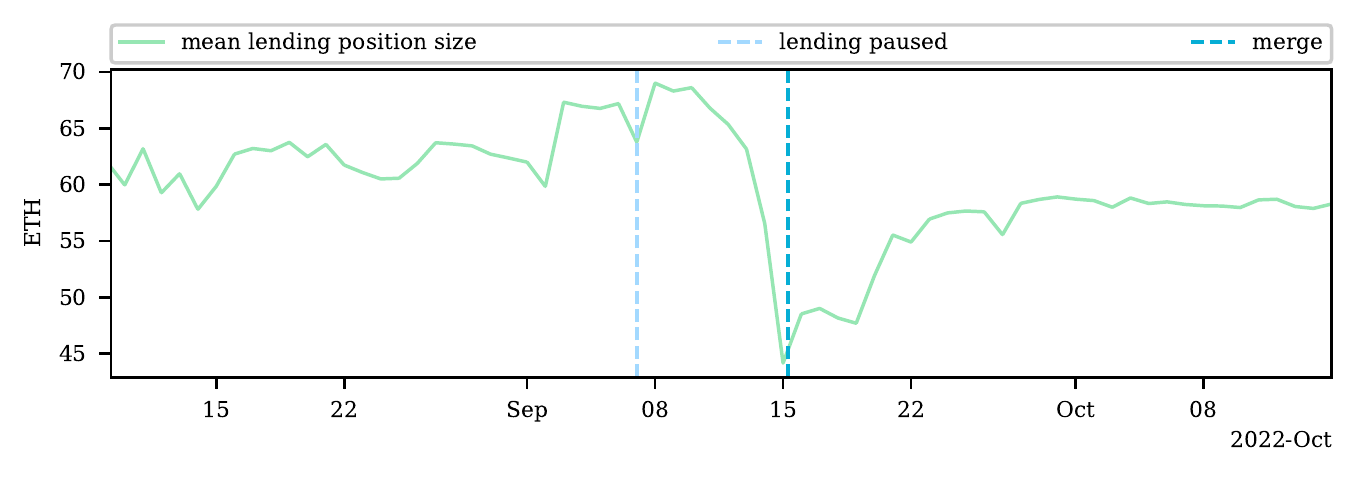}\vspace{-2pt}
\caption{Mean size of the lending positions on AAVE as a function of time. Notice the drop prior to the merge, indicating that primarily large lenders exited, while smaller liquidity providers were left back.} \label{fig:average_LP}\vspace{-2pt}
\end{figure}

Figure~\ref{fig:average_LP} visualizes the average size of a lending position over time. Observe the clear drop prior to the merge indicating that primarily larger liquidity providers exited, whereas the smaller players were more likely to stay put. We presume that the larger and likely more sophisticated liquidity providers exited the AAVE with their ETH in time, while the smaller and likely less sophisticated lender remained stuck in the pool once the utilization reached 100\%. Thus, the smaller lenders bore the brunt of the losses as they missed out on the hard-fork arbitrage. 

\begin{figure}[t]
\centering
\includegraphics[scale=1,right]{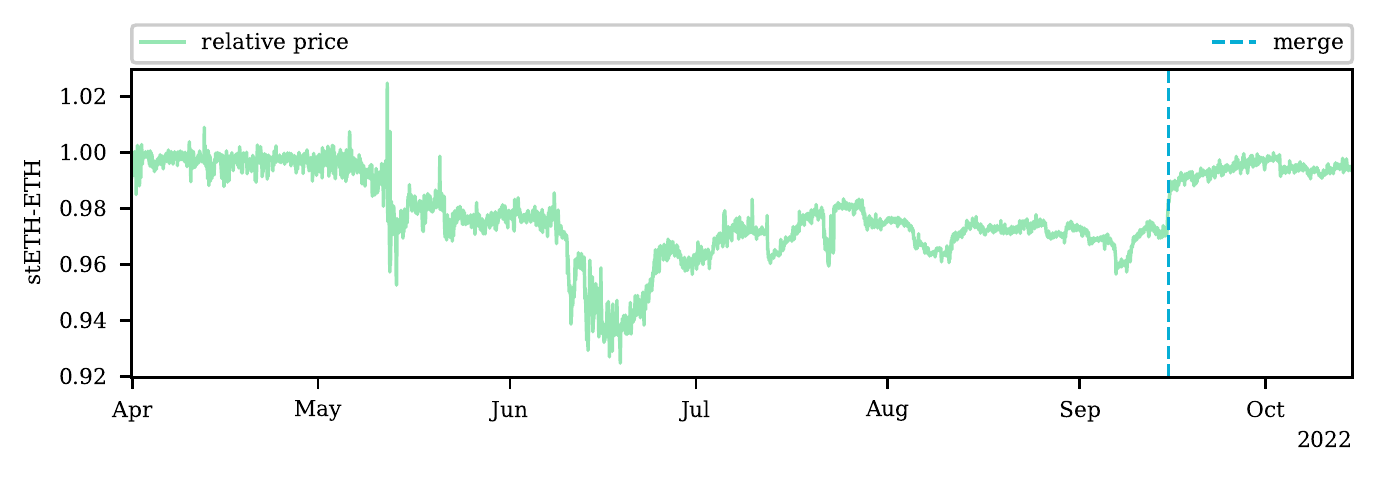}\vspace{-2pt}
\caption{The stETH price over time. In the lead-up to the merge stETH was trading at a significant discount but returned to price parity after the merge.} \label{fig:rel_prices_stETH}\vspace{-2pt}
\end{figure}

Unlike Compound, AAVE allows stETH to be used as collateral. The value of stETH stems from the staking rewards as well as the fact that in the future stETH holders will be able to swap stETH for ETH. However, unlike current ETH holders, stETH owners did not receive ETHW after the merge. Therefore, unsurprisingly, stETH was trading at a discount to ETH in the months before the merge (cf. Figure~\ref{fig:rel_prices_stETH}). Observe that this discount is comparable to the value of ETHW. After the merge, the stETH-ETH price recovered and was again trading close to parity. 

\begin{figure}[t]
\centering
\includegraphics[scale=1,right]{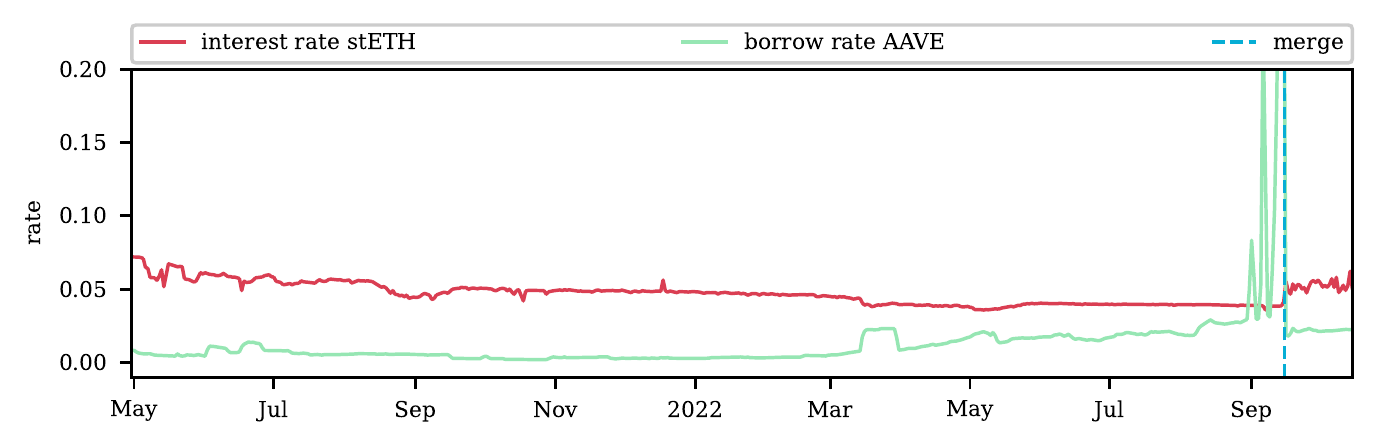}\vspace{-2pt}
\caption{Annualized stETH staking rewards (red) and the ETH borrowing rate (green) on AAVE. In the months prior to the merge the staking rewards were significantly higher than the borrowing costs, making it profitable for traders to use leverage to buy stETH.} \label{fig:rates_stETH}\vspace{-2pt}
\end{figure}

While AAVE does not facilitate stETH borrowing, users can use stETH as collateral to borrow ETH,  which in turn can be used to acquire yet more stETH. AAVE allowed 1 stETH to be used to borrow 0.73 ETH -- enabling a popular trading strategy as the difference between the staking rewards and the ETH borrowing costs was quite significant (cf. Figure~\ref{fig:rates_stETH}). For months the staking rewards were significantly higher, enticing investors to take on leverage. However, this reversed in the lead-up to the merge as borrowing rates spiked. This is significant as we find 20.3\% of the stETH market capitalization is deposited on AAVE (cf. Appendix~\ref{app:steth}). As AAVE does not support stETH borrowing, there is no reason to deposit other than to take out loans. More than one-fifth of the stETH pledged as collateral makes mass liquidations a real threat to the stETH price and, thereby, to the blockchain consensus layer, as staking power can be acquired at a discount.

\section{Discussion}
\TA{Intervention by DAOs.} While borrowing costs in the days prior to the merge were high relative to the rates typically seen on these protocols, they were still far lower than the payoff an ETH borrower could expect. Thus, rates were too low from the lender's perspective. Lenders were neither adequately compensated for forgoing this arbitrage opportunity nor for the uncertainty that surrounded DeFi in general and lending protocols in particular. A borrower wishing to borrow for the last block prior to the merge should have paid interest at least equal to the price of one ETHW token --- orders of magnitudes more than the effective rate. Only such a rate would have compensated a lender for not profiting from the hard-fork-arbitrage. Furthermore, AAVE liquidity providers were unable to withdraw their funds as the utilization rate approached 100\%, thus depriving them of even having the option to withdraw. Ultimately, the profits the arbitrageurs made at the expense of liquidity providers, whom the protocols failed to adequately compensate. 

In comparison, the largest centralized crypto exchange, Binance, encouraged users to repay their ETH and, furthermore, withheld the ETHW that was awarded for borrowed ETH~\cite{binancemerge}. Whether this course of action was more equitable is surely debatable. In fact, Binance suspended withdrawals altogether~\cite{binancemerge}. However, it avoided the situation of charging extremely high borrowing rates but still kept a lending market open. This was particularly beneficial for traders who wanted to borrow ETH for reasons other than speculating on the ETHW tokens.  

This discussion highlights that the merge was an extraordinary situation that led to interventions on both decentralized and centralized lending platforms. They capped borrowing, limited rates, or tried to deter speculators by withholding the ETHW tokens. For DeFi lending protocols, this meant having to give up their `no intervention' mantra. While these interventions safeguarded the protocols as a whole, they were paid for by the liquidity providers. This lack of compensation contributed to the DeFi lending market drying up. Given the continued rapid growth of DeFi and the particular importance these protocols play in this ever-more intertwined system, the lack of liquidity in such extraordinary market situations poses a grave threat to the broader Ethereum ecosystem. As shown during the 2008 financial crisis, the drying up of the lending market can greatly exacerbate market downturns. 

Furthermore, we stress that the merge was announced well in advance leaving ample time for the protocols to come up with and implement their proposals. While our study focuses on one particular hard-fork, Ethereum has gone through more than a dozen hard-forks since its genesis~\cite{2023EthereumFork}. Not all of these were announced far ahead of time. Thus, future more grave outcomes for lending protocols in the face of hard forks cannot be discounted. Additionally, we note that external market shocks are rarely as foreseeable as in this case. Rapid, unexpected developments could deprive DAOs of this course of action and pose greater threats to the viability and security of the protocol. For instance, the recent accumulation of bad debt on AAVE left by an attacker that borrowed a large amount of CRV tokens for short-selling, could not be prevented due to events unfolding much more rapidly~\cite{eigenphi2023how}. 

\T{Security concerns beyond lending protocols.} We note the potential ramifications of the highly leveraged stETH positions the borrowing spirals created. The reversal of the difference between staking rewards and borrowing costs made these leveraged positions in stETH unprofitable. This posed a grave security threat, as a total of 20.3\% of stETH was locked on AAVE. Liquidations due to rising borrowing costs and/or a falling stETH price would further devalue the collateral of other leveraged stETH holders, resulting in a downward spiral. The ramifications thereof would spill over to the wider DeFi ecosystem, as users could, for example, acquire stETH and its staking power at a significant discount. As LIDO accounts for more than a fourth of staking power on the Beacon chain~\cite{2023beaconcha} and given the size of these lending protocols, their viability is crucial to DeFi as a whole.

\T{Market inefficiencies.} Finally, we add that the ETH lending market on AAVE and Compound is an example of market inefficiencies in the cryptocurrency market. In an efficient market, i.e., a market where prices reflect all relevant information~\cite{malkiel1989stock}, the hard-fork-arbitrage we studied in this work should not exist as the combined market value of ETH and ETHW after the fork should be equal to the market value of ETH before the fork (no arbitrage condition). Thus, the hard-fork-arbitrage is an empirical example of market inefficiencies in DeFi. 

\section{Conclusions}
Given the central role of lending protocols in DeFi and the composability of the latter, the stability of these protocols is crucial to the entire ecosystem. Therefore, unsurprisingly, concern grew in the months leading up to the merge that hard-fork-arbitrageurs would borrow large amounts of ETH, which would drive up rates and potentially lead to mass liquidation. Both Compound and AAVE saw no alternative to intervention and effectively capped borrowing. 

Our analysis finds that these interventions may have helped prevent widespread liquidations. However, these interventions led to market distortions and were made at the expense of the protocol's liquidity providers. On the other hand, large borrowers like Alameda Research, who speculated on the hard-fork-arbitrage transferred proceeds from the hard-fork arbitrage worth more than 13~Mio~US\$ at the time of the merge to centralized exchanges. These tokens were in effect extracted from the liquidity providers, who were by far not fairly compensated for either their service or for the risk they bore. Furthermore, as the utilization rate approached 100\%, the lending market ceased to function. Neither could liquidity providers withdraw nor could new debt be taken on, effectively drying up the DeFi lending market.

Finally, we find that the increased complexity resulting from the ever-increasing composability of DeFi poses security concerns not only for DeFi protocols but even for the consensus layer. For example, over one-fifth of the ETH staked through LIDO was locked on AAVE as collateral during the merge. Widespread liquidations would have led to a dramatic drop in the price of stETH, effectively giving a discount to anyone wishing to acquire staking power.
\clearpage

\newpage
\bibliography{references}
\newpage


\appendix
\section{stETH Market Capitalization and AAVE stETH Collateral}\label{app:steth}
\begin{figure}[ht]
\centering
\includegraphics[scale=1,right]{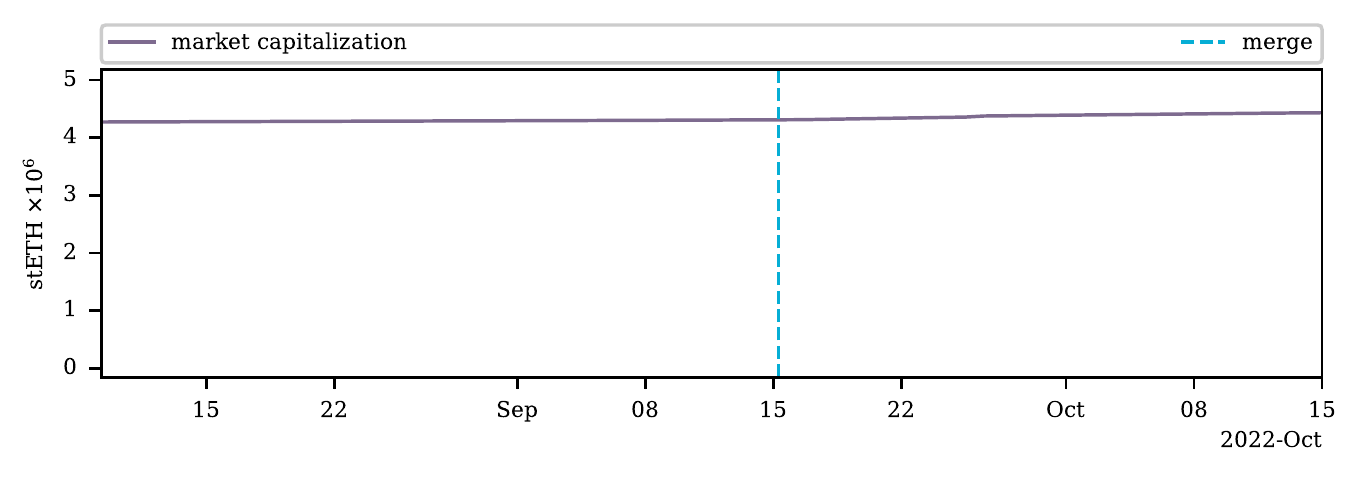}\vspace{-15pt}
\caption{Market capitalization of stETH.} \label{fig:stETH_total}\vspace{-25pt}
\end{figure}

We plot the market capitalization of stETH before and after the merge in Figure~\ref{fig:stETH_total}. The stETH market capitalization corresponds to the combined ETH balance of LIDO validators on the Beacon chain. We point out that the stETH market capitalization is ever-increasing in that time frame, as withdrawals from the Beacon chain have not been activated. At the same time additional ETH is staked on the Beacon chain and the staked ETH balance increases as the validators are receiving rewards for performing their duties. Note that validators that do not carry out their duties correctly will be slashed and the ETH balance will reduce, this, however, did not happen to a large extent for LIDO validators. 

\begin{figure}[h]
\centering
\includegraphics[scale=1,right]{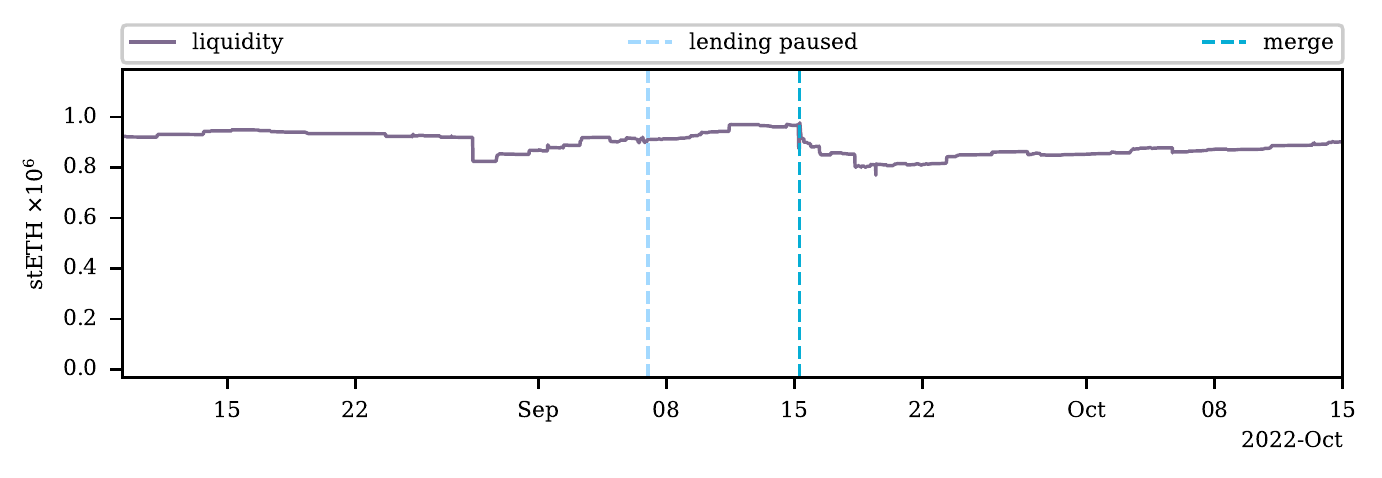}\vspace{-15pt}
\caption{Amount of stETH locked on AAVE around the merge.} \label{fig:stETH_liquidity_aave}\vspace{-10pt}
\end{figure}

We also plot the amount of stETH locked on AAVE in Figure~\ref{fig:stETH_liquidity_aave}. As stETH cannot be borrowed, the stETH on AAVE is likely used as collateral to take out debt. A popular strategy for stETH holders was to take out ETH debt to be able to stake additional ETH with LIDO~\cite{Primoz2022arc}. Thus, it is both astonishing and worrying at the same time that around 20\% of all stETH are locked on AAVE. A price drop of stETH cloud cause liquidations of loans on AAVE with stETH collateral which would further apply downward pressure on the stETH price. Thus, the high levels of stETH locked on AAVE pose a security concern for the Ethereum consensus layer.

\section{Cumulative Rates on AAVE and Compound}\label{app:cumulative}
\begin{figure}[t]
\centering
\includegraphics[scale=1,right]{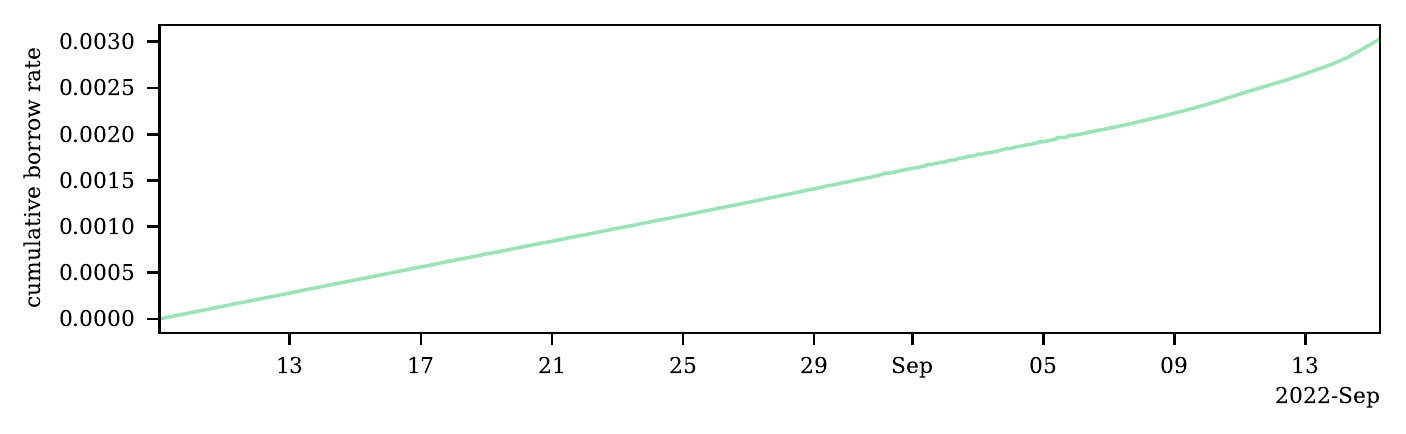}\vspace{-12pt}
\caption{The cumulative interest rate for a borrower on Compound starting from 9 August 2022 until the merge.} \label{fig:cumulative_compound}\vspace{-12pt}
\end{figure}

We plot the cumulative borrowing rate a borrower would have paid for an ETH debt taken out on 9 August 2022, the day ETHW started trading, and held until the merge, 15 September 2022. In Figure~\ref{fig:cumulative_compound}, we show the cumulative rate that would have been paid by an ETH borrower. A borrower would have paid around 0.03\% for an ETH debt held for that time window -- significantly less than the relative value of ETHW compared to ETH during the merge. Notice that the rate increases almost linearly as a consequence of the relatively low borrowing rate on Compound in the lead-up to the merge. 

\begin{figure}[t]
\centering
\includegraphics[scale=1,right]{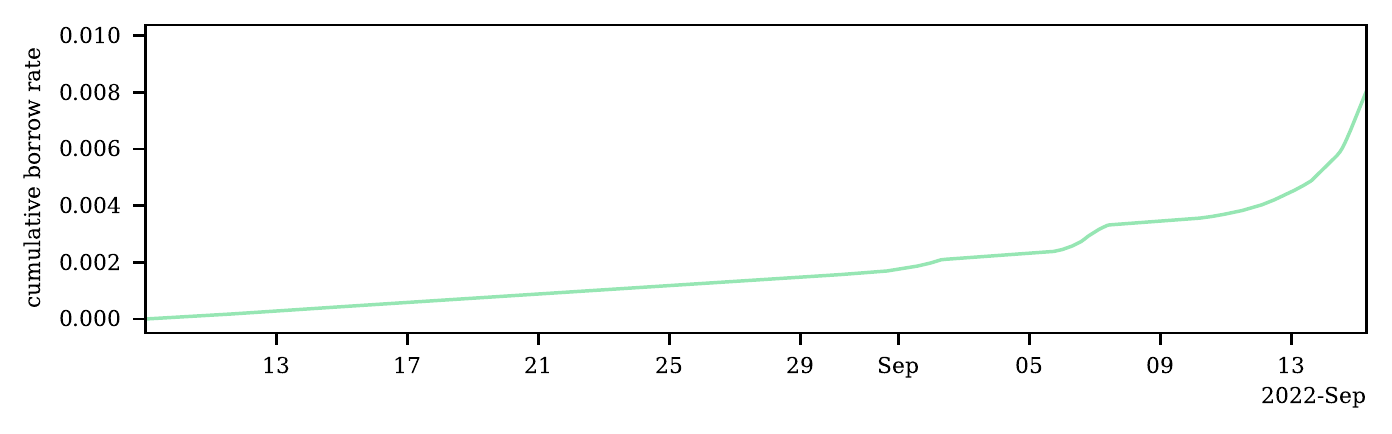}\vspace{-12pt}
\caption{The cumulative interest rate for a borrower on AAVE starting from 9 August 2022 until the merge.} \label{fig:cumulative_aave}\vspace{-10pt}
\end{figure}

In Figure~\ref{fig:cumulative_aave}, on the other hand, we plot the cumulative borrowing rate for an ETH borrower on AAVE during the same time period. The cumulative rate paid on AAVE would have been higher than on Compound with 1\%, but still significantly less than the price of ETHW in terms of ETH during the merge. The cumulative borrowing rate on AAVE rapidly increased starting from 7 September 2022 as a result of the sharp increase in the borrowing rate ahead of the merge.

\section{Interest Rate Curves}\label{app:interestrate}

\begin{figure}[ht]
\centering
\includegraphics[scale=1,right]{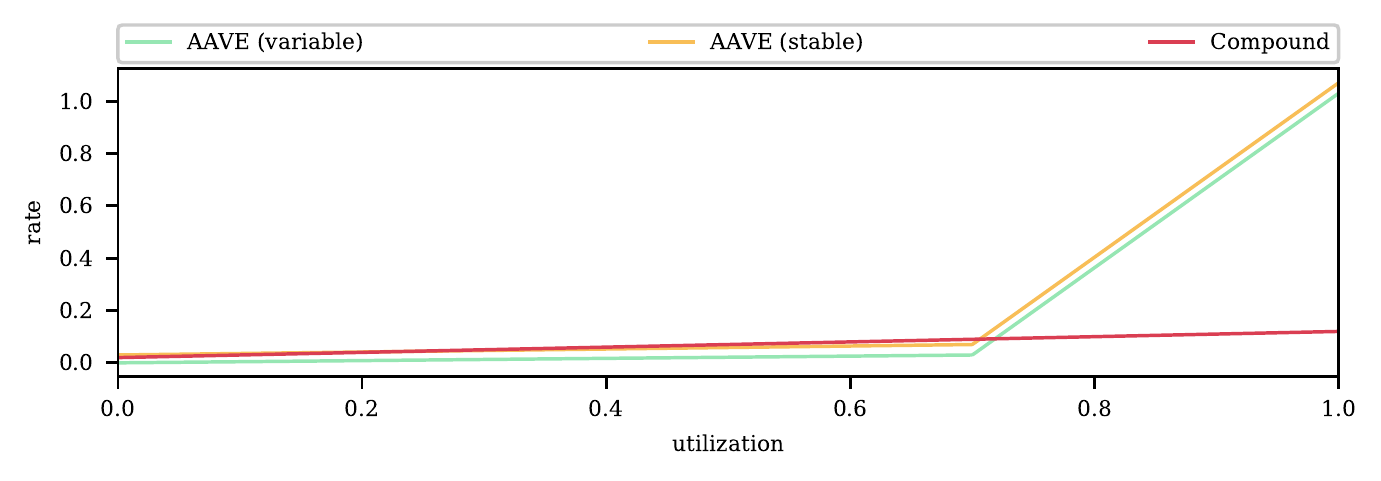}\vspace{-8pt}
\caption{ETH borrowing rates on AAVE and Compound as a function of the utilization.}\label{fig:ratesutilization}\vspace{-10pt}
\end{figure}

The interest rate for the two lending protocols are given in Section~\ref{sec:lendingprotocol}. Here, we show the interest rate curves as a function of utilization for AAVE and Compound in Figure~\ref{fig:ratesutilization}. Furthermore, as described in Section~\ref{sec:compound}, Compound updated their interest rate model in anticipation of the merge to the jump interest rate model that qualitatively looks similar to that of AAVE.

\end{document}